\documentclass[sigconf]{acmart}

\def\BibTeX{{\rm B\kern-.05em{\sc i\kern-.025em b}\kern-.08emT\kern-.1667em\lower.7ex\hbox{E}\kern-.125emX}}

\usepackage[utf8]{inputenc}
\usepackage{balance}
\usepackage{booktabs}
\usepackage{graphicx}
\usepackage{array}
\usepackage{algpseudocode}
\usepackage{amsfonts}
\usepackage{amsmath}
\usepackage{amssymb}
\usepackage{mathtools}
\usepackage{multirow,bigdelim,dcolumn}
\usepackage{subscript}
\usepackage{color, colortbl, soul}
\usepackage{subcaption}
\usepackage{comment}
\usepackage{tikz}
\usetikzlibrary{matrix, decorations.pathreplacing, calc, positioning, arrows}

\copyrightyear{2019}
\acmYear{2019}
\setcopyright{acmlicensed}
\acmConference[WebSci 2019]{ACM Web Science Conference 2019}{June 30--July 3, 2019}{Boston, MA, USA}
\acmPrice{}
\acmDOI{}
\acmISBN{}

\begin{document}

\title[RTBust: Exploiting Temporal Patterns for Botnet Detection on Twitter]{\textsc{RTbust}: Exploiting Temporal Patterns\\for Botnet Detection on Twitter}

\author{Michele Mazza}
\affiliation{
  \institution{IIT-CNR, Italy}
}
\email{michele.mazza@iit.cnr.it}

\author{Stefano Cresci}
\authornote{This is the corresponding author.}
\orcid{0000-0003-0170-2445}
\affiliation{  \institution{IIT-CNR, Italy}
}
\email{stefano.cresci@iit.cnr.it}

\author{Marco Avvenuti}
\affiliation{
  \institution{University of Pisa, Italy}
}
\email{marco.avvenuti@unipi.it}

\author{Walter Quattrociocchi}
\affiliation{
  \institution{Ca' Foscari University of Venice, Italy}
}
\email{w.quattrociocchi@unive.it}

\author{Maurizio Tesconi}
\affiliation{
  \institution{IIT-CNR, Italy}
}
\email{maurizio.tesconi@iit.cnr.it}

\renewcommand{\shortauthors}{M. Mazza \textit{et al.}}

\begin{abstract}
Within OSNs, many of our supposedly online friends may instead be fake accounts called \textit{social bots}, part of large groups that purposely re-share targeted content. Here, we study retweeting behaviors on Twitter, with the ultimate goal of detecting retweeting social bots. We collect a dataset of 10M retweets. We design a novel visualization that we leverage to highlight benign and malicious patterns of retweeting activity. In this way, we uncover a ``normal'' retweeting pattern that is peculiar of human-operated accounts, and 3 suspicious patterns related to bot activities. Then, we propose a bot detection technique that stems from the previous exploration of retweeting behaviors. Our technique, called \textsc{Retweet-Buster} (\textsc{RTbust}), leverages unsupervised feature extraction and clustering. An LSTM autoencoder converts the retweet time series into compact and informative latent feature vectors, which are then clustered with a hierarchical density-based algorithm. Accounts belonging to large clusters characterized by malicious retweeting patterns are labeled as bots. \textsc{RTbust} obtains excellent detection results, with $F1 = 0.87$, whereas competitors achieve $F1 \leq 0.76$. Finally, we apply \textsc{RTbust} to a large dataset of retweets, uncovering 2 previously unknown active botnets with hundreds of accounts.
\end{abstract}

\keywords{Social bots, retweet patterns, OSN security, Twitter}

\maketitle

\makeatletter{}\section{Introduction}
\label{sec:introduction}
In 2016 the Oxford dictionary elected "Post-Truth" as the word of the year and in 2017 Collins dictionary did the same for "Fake News".
In 2017 the World Economic Forum raised a warning on the potential distortion effect of social media on user perceptions of reality\footnote{\scriptsize\url{http://reports.weforum.org/global-risks-2017/acknowledgements/}}.
Recent studies, targeting Facebook, showed the tendency of the users to interact with information adhering to their preferred narrative~\cite{del2016spreading,bessi2015science} and to ignore dissenting information~\cite{zollo2017debunking}. 
Confirmation bias seems to account for user decisions about consuming and spreading content and at the same time, aggregation of favored information within those communities reinforces selective exposure and group polarization~\cite{sunstein2002law,quattrociocchi2017inside}.
As we interact with our peers, we are exposed to the effects of echo chambers, which may result in polarization and hate speech~\cite{del2016echo}.
In this scenario the role of bots is not clear, both for the difficulties in quantifying their impact as well as for the accuracy of detection algorithms. However, any of our supposedly online friends may instead be fake, automated accounts (\textit{social bots}), and a part of large coordinated groups that purposely (re-)share targeted content~\cite{ferrara2016}. Along this challenge, in this work we propose a new method for the detection of groups of bots, accounting for their statistical traces induced by their coordinated behavior on Twitter.

Specifically, we focus on re-sharing (i.e., retweeting) patterns with the ultimate goal of detecting retweeting social bots. Indeed, by artificially boosting retweet counts, these bots may affect users' popularity or influence, thus ``reshaping political debates~\cite{bovet2019influence,stella2018bots}. They can defraud businesses and ruin reputations"\footnote{\scriptsize\url{https://www.nytimes.com/interactive/2018/01/27/technology/social-media-bots.html}}. Moreover, large retweet counts indicate influential users, and can be monetized. Thus, there are strong economic and sociopolitical incentives for malicious bots to tamper with retweets~\cite{giatsoglou2015retweeting,steward2018}. Hence, fast and accurate removal of these bots might be crucial for ensuring the healthiness of our online social ecosystems.

Both researchers and OSN administrators have recently been very active towards the detection of social bots, and many different techniques have been proposed to this end. Unfortunately, the same also applies to malicious bot developers. In fact, as soon as new detection techniques are deployed, bot developers tweak the characteristics of their accounts, thus allowing them to evade detection~\cite{Cresci2017}. This ``never-ending clash'' led to the current situation where social bots are so sophisticatedly engineered as to mimic legitimate accounts, becoming almost indistinguishable from them~\cite{cresci2018proaction}. A straightforward consequence of this situation is that standard machine learning classification approaches, where each account is analyzed individually, are no longer profitable. Instead, the scientific frontier of bot detection now focuses on groups of suspicious accounts as a whole. Analyzing groups has the advantage that, no matter how sophisticated a single bot can be, a large enough group of them will still leave traces of automation, since they do share a common goal (e.g., increasing someone’s popularity)~\cite{Cresci2017}. For the same reasons, unsupervised approaches are preferred over supervised ones~\cite{yang2019arming}.

The possibility to exploit more data for the analysis, opened up by the approaches that target groups rather than individual accounts, is however counterbalanced by the difficulties in collecting and processing that much data.
For instance, the behavior-based technique described in~\cite{cresci2016dna,cresci2017tdsc} requires collecting and comparing data of the Twitter timelines of all analyzed accounts. Similarly, graph-based techniques such as~\cite{yang2014uncovering,jiang2016} require building and analyzing the social graph of a large group of accounts. The amount of data and computational power needed to complete these analyses, inevitably limits the large-scale applicability of these techniques. In short, the next generation of bot detection techniques should strike the balance between accuracy and data/algorithmic efficiency.

\textbf{Contributions.} We propose a novel technique for the detection of retweeting social bots, called \textsc{Retweet-Buster} (\textsc{RTbust}), having all the desirable features previously discussed. Our technique only requires the timestamps of retweets (and of the retweeted tweets) for each analyzed account, thus avoiding the need for full user timelines or social graphs. Then, it compares the temporal patterns of retweeting activity of large groups of users. Leveraging previous findings in the field, \textsc{RTbust} looks for groups of accounts with distinctive and synchronized patterns. Evaluation results on a large dataset of retweets demonstrate excellent bot detection performances, with $F1 = 0.87$, whereas competitors achieve $F1 \leq 0.76$. Summarizing, our detailed contributions are as follows.
\begin{itemize}
    \item We analyze retweeting behaviors of a large set of users by introducing a simple -- yet effective -- visualization, which we then leverage to highlight benign and malicious patterns of retweeting activity.
    \item We design a group-analysis technique that is capable of detecting accounts having the same retweeting patterns. Accounts belonging to a large group characterized by the same malicious patterns are labeled as bots.
    \item We compare detection results of our technique with those obtained by baselines and other state-of-the-art techniques, demonstrating the effectiveness of our approach.
    \item By applying our technique to a large dataset of retweets, we uncover 2 previously unknown active botnets.
\end{itemize}

\makeatletter{}\section{Related Work in bot detection}
\label{sec:relwork}
The vast majority of previous attempts at bot detection are based on supervised machine learning~\cite{Cresci2017}. The first challenge in developing a supervised detector is related to the availability of a ground truth (i.e., labeled) dataset, to be used in the learning phase of the classifier. In most cases, a real ground truth is lacking and the labels are simply given by human operators that manually analyze the data. Critical issues arise since there is no ``standard'' definition of what a social bot is~\cite{maus2017typology,yang2019arming}. Moreover, humans have been proven to largely fail at spotting sophisticated bots, with only $\simeq 24\%$ bots correctly labeled as such~\cite{Cresci2017}. As anticipated, these criticalities support the development of unsupervised techniques.

Regarding features to exploit for the detection, 3 classes have been mainly considered: (i) profile features~\cite{yang2013,davis2016,kater2016you}; (ii) features extracted from the posts, such as posting behavior and content of posted messages~\cite{lee2014early,miller2014,chavoshi2016debot}; and (iii) features derived from the social or interaction graph of the accounts~\cite{yang2014uncovering,jiang2016,jiang2016inferring,liu2017holoscope}. The classes of features exploited by the detection technique have a strong impact on both the performances of the detector as well as its efficiency~\cite{cresci2015}. For instance, in Twitter it has been demonstrated that those features that mostly contribute towards the predictive power of bot detectors (e.g., measures of centrality in the social graph), are also the most costly ones.

The difficulties in detecting sophisticated bots with supervised approaches that are based on the analysis of individual accounts, recently gave rise to a new research trend that aims to analyze groups of accounts as a whole~\cite{Cresci2017}. This new approach to bot detection is proving particularly effective at detecting coordinated and synchronized bots, such as those targeted in our work. For instance, the technique discussed in~\cite{cresci2016dna,cresci2017tdsc} associates each account to a sequence of characters that encodes its behavioral information. Such sequences are then compared between one another to find anomalous similarities among sequences of a subgroup of accounts. The similarity is computed by measuring the longest common subsequence shared by all the accounts of the group. Accounts that share a suspiciously long subsequence are then labeled as bots. Instead, the family of systems described in~\cite{jiang2016,liu2017holoscope,liu2018contrast} build a bipartite graph of accounts and their interactions with content (e.g., retweets to some other tweets) or with other accounts (e.g., becoming followers of other accounts). Then, they aim to detect anomalously dense blocks in the graph, which might be representative of coordinated and synchronized attacks. A possible drawback of group approaches is that they exacerbate challenges related to data and algorithmic costs, since they typically involve a large number of comparisons between all accounts in a given group~\cite{vo2017revealing,cresci2017tdsc}.

Lastly, a trailblazing direction of research involves the application of adversarial machine learning to bot detection. Until now, bot detection has mostly followed a \textit{reactive} schema, where countermeasures are taken only after having witnessed evidence of bot mischiefs~\cite{cresci2018proaction}. Instead, in~\cite{cresci2018proaction} is proposed an adversarial framework enabling \textit{proactive} analyses. The framework has been later instantiated in~\cite{cresci2019capability} by using evolutionary algorithms to test current detection techniques against a wider range of unseen, sophisticated bots. Another preliminary work in adversarial bot detection is described in~\cite{grimme2018changing}. Among the positive outcomes of adversarial approaches to bot detection, is a more rapid understanding of the drawbacks of current detectors and the opportunity to gain insights into new features for achieving more robust and more reliable detectors. 
\makeatletter{}\begin{figure*}[t]
	\begin{minipage}[b]{0.225\textwidth}
        \centering
        \includegraphics[width=\textwidth]{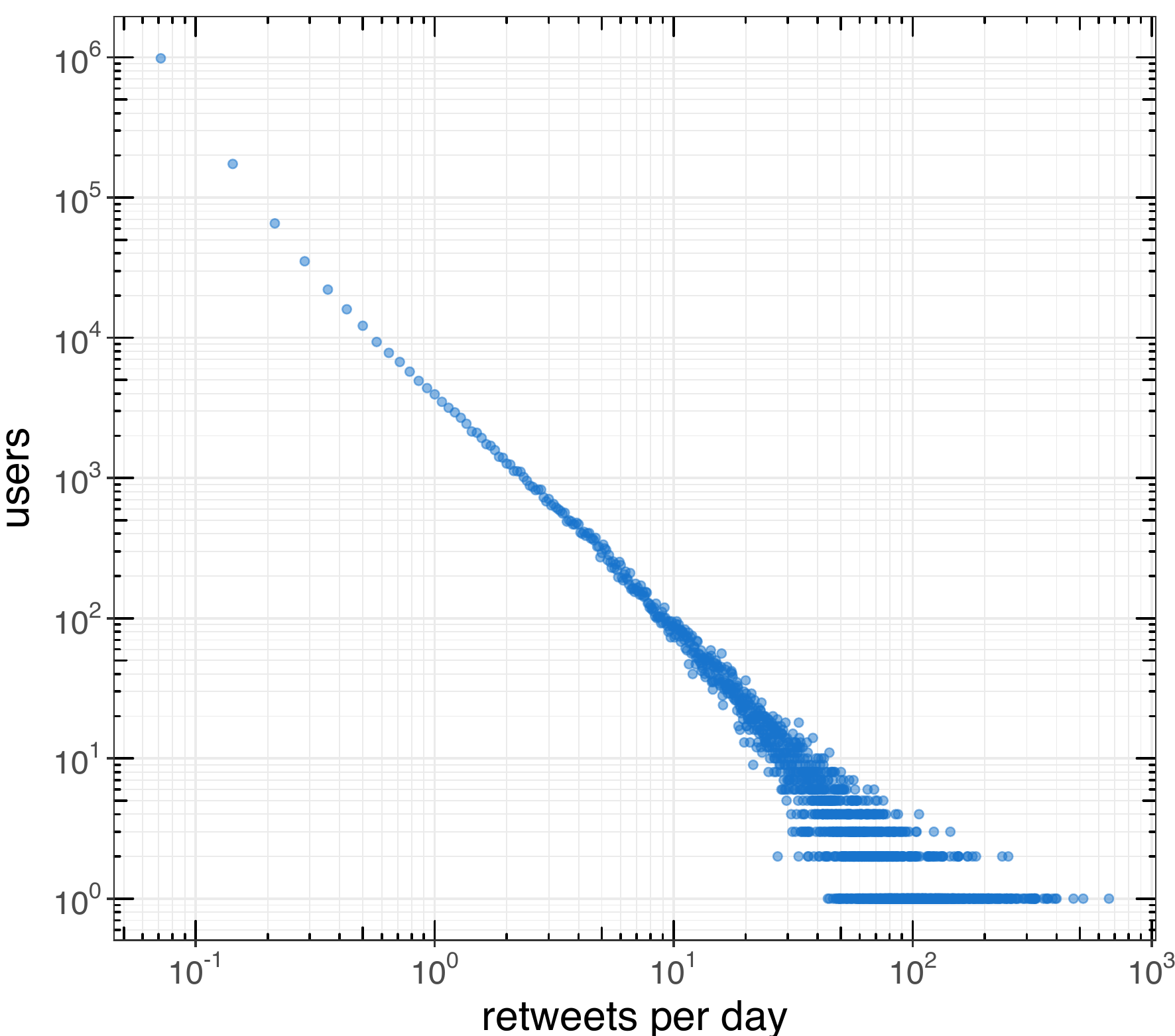}
        \caption{Daily retweets per user.\label{fig:retweets-per-user}}
	\end{minipage}
	\hspace{0.04\textwidth}
	\begin{minipage}[b]{0.225\textwidth}
        \centering
        \includegraphics[width=\textwidth]{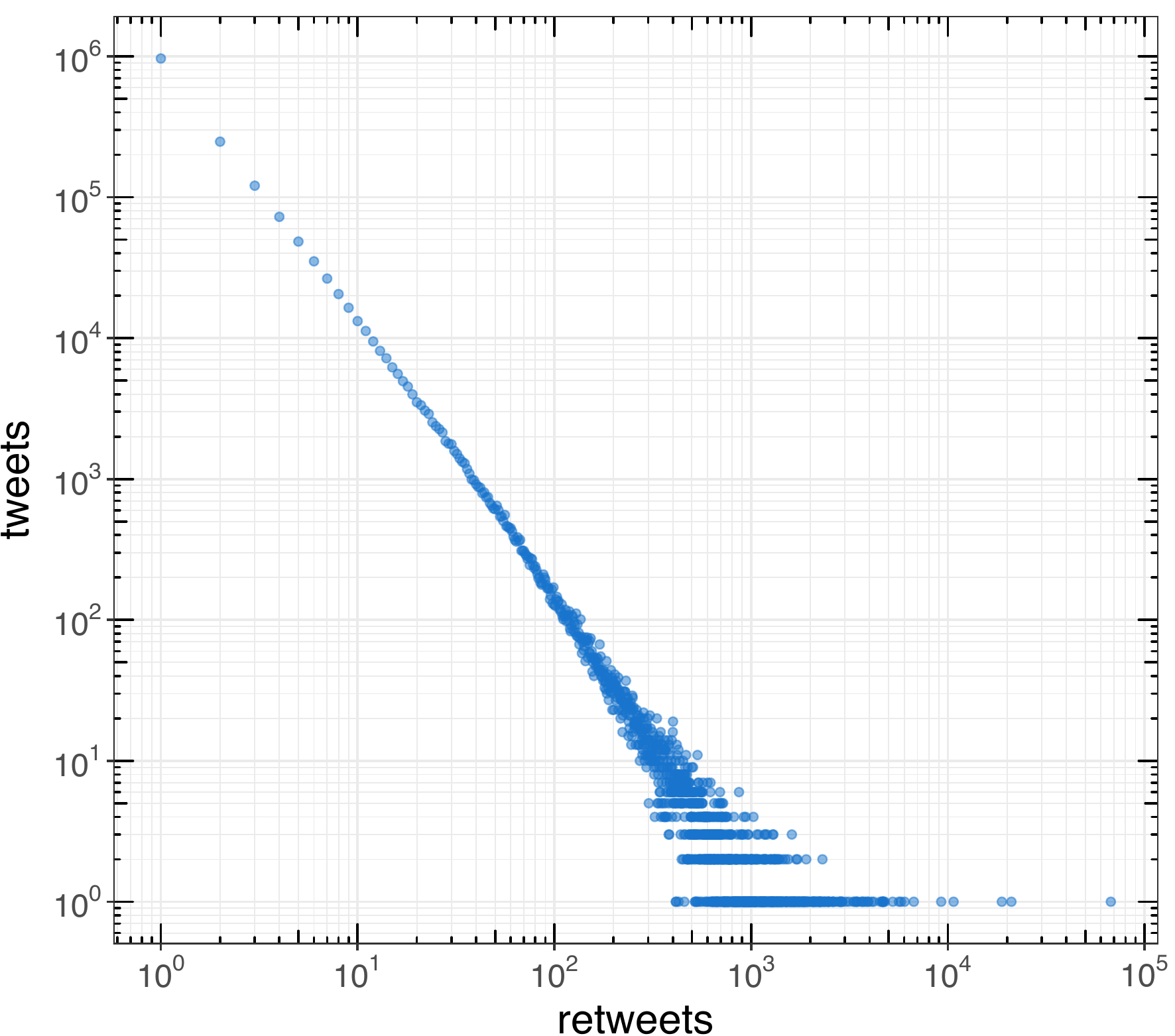}
        \caption{Total retweets per tweet.\label{fig:retweets-per-tweet}}
	\end{minipage}
	\hspace{0.04\textwidth}
	\begin{minipage}[b]{0.375\textwidth}
        \centering
        \includegraphics[width=\textwidth]{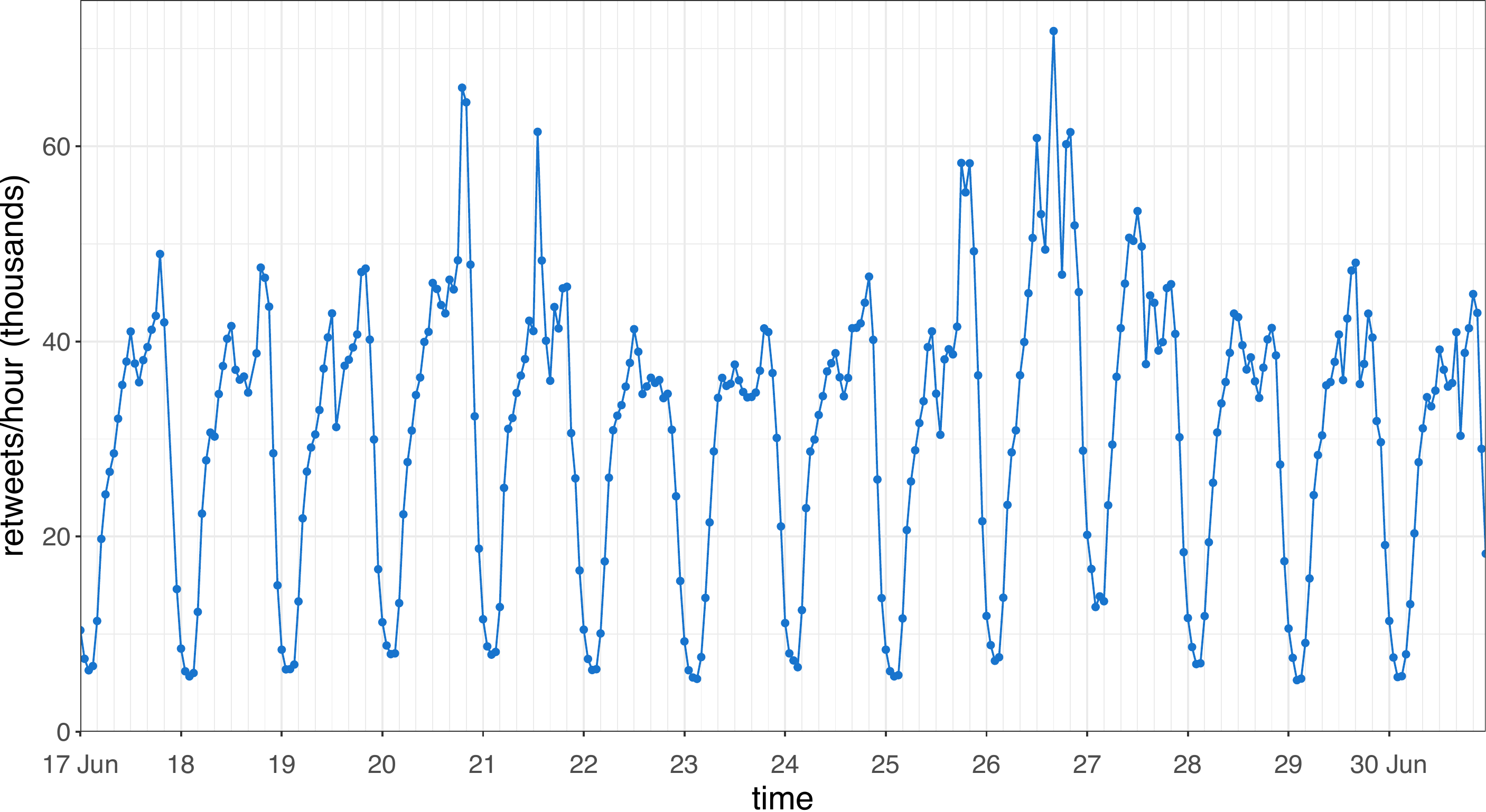}
        \caption{Hourly volume of retweets across the 2 considered weeks.\label{fig:retweets-timeserie}}
	\end{minipage}
\end{figure*}

\section{Data collection and annotation}
\label{sec:dataset}
Our dataset for this study is composed of \textit{all} Italian retweets shared in a 2 weeks time span -- specifically, between 17 and 30 June, 2018 (inclusive). Overall, our dataset comprises 9,989,819 retweets, shared by 1,446,250 distinct users. Thus, on average, each user in our dataset retweeted 7 times per day, in line with recent statistics reporting between 2 and 50 daily retweets for legitimate users~\cite{firdaus2018retweet}. However, our dataset also includes many ``extreme'' users, as visible in Figure~\ref{fig:retweets-per-user} showing the distribution of retweets per day per user, which features a typical heavy-tailed shape. The 9,989,819 retweets are related to 1,691,865 distinct original tweets, which are also included in the dataset. Figure~\ref{fig:retweets-per-tweet} shows the distribution of retweets per original tweet, while Figure~\ref{fig:retweets-timeserie} shows the hourly volume of retweets across the 2 considered weeks.

For each tweet, retweet, and user in our dataset we have access to all Twitter metadata fields, as provided by Twitter APIs. To collect this dataset, we resorted to Twitter Premium Search API\footnote{\scriptsize\url{https://developer.twitter.com/en/docs/tweets/search/api-reference/premium-search.html}} with the following query parameters\footnote{\scriptsize\url{https://developer.twitter.com/en/docs/tweets/search/overview/premium\#AvailableOperators}}: \texttt{lang:IT} and \texttt{is:retweet}. Here, the exploitation of the Premium Search API is important since it allowed us to build a \textit{complete} dataset of retweets. In fact, the Standard Search API\footnote{\scriptsize\url{https://developer.twitter.com/en/docs/tweets/search/api-reference/get-search-tweets.html}} used in the majority of previous works, does not guarantee completeness, meaning that not all tweets matching the query criteria are returned. Notably, although our dataset for this study is limited to tweets in the Italian language, both the data collection approach and the analytical process described in the remainder of the paper, are totally language independent. The language of collected tweets can be easily changed with the \texttt{lang:} parameter, and our analyses only exploit timestamps.

After data collection, we performed data filtering and annotation. A manual inspection of those users exhibiting the largest number of retweets per day quickly revealed their automated nature. However, it turned out that all such accounts, despite being bots, were not malicious ones. Their automated nature was manifest, they did not try to disguise as human-operated accounts, and they were not acting coordinatedly nor trying to inflate the popularity of some specific content. The presence of this kind of bots is well known~\cite{gilani2017classification}. They do not pose a threat to OSNs and, in fact, some of them are even benign~\cite{savage2016botivist,avvenuti2017hybrid}. Thus, we excluded such accounts from future analyses, since they are not the target of our work. Similarly, we also excluded accounts featuring a very small number of retweets. Operationally, we set our filtering thresholds by leveraging statistics in~\cite{firdaus2018retweet} -- that is, we retained only those users with a mean number of retweets per day $\geq 2$ and $\leq 50$. In this way, we ended up with 63,762 distinct users exhibiting human-like retweeting behaviors. The goal of our next analyses is to tell apart the sophisticated human-like bots from the real human-operated accounts.

Although our detection technique is unsupervised and hence it does not require a labeled ground truth, we nonetheless carried out manual annotation of a small subset of our dataset. This is useful in order to evaluate the extent to which our technique is capable of correctly spotting the bots. We thus annotated $\simeq 1,000$ accounts from our dataset, following the latest annotation guidelines for datasets containing social bots~\cite{Cresci2017}. We ended up with an almost balanced annotated dataset, comprising $51\%$ bots and $49\%$ legitimate (i.e., human-operated) accounts. 
\makeatletter{}\begin{figure*}[t]
	\begin{subfigure}[b]{0.25\textwidth}
        \centering
        \includegraphics[width=\textwidth]{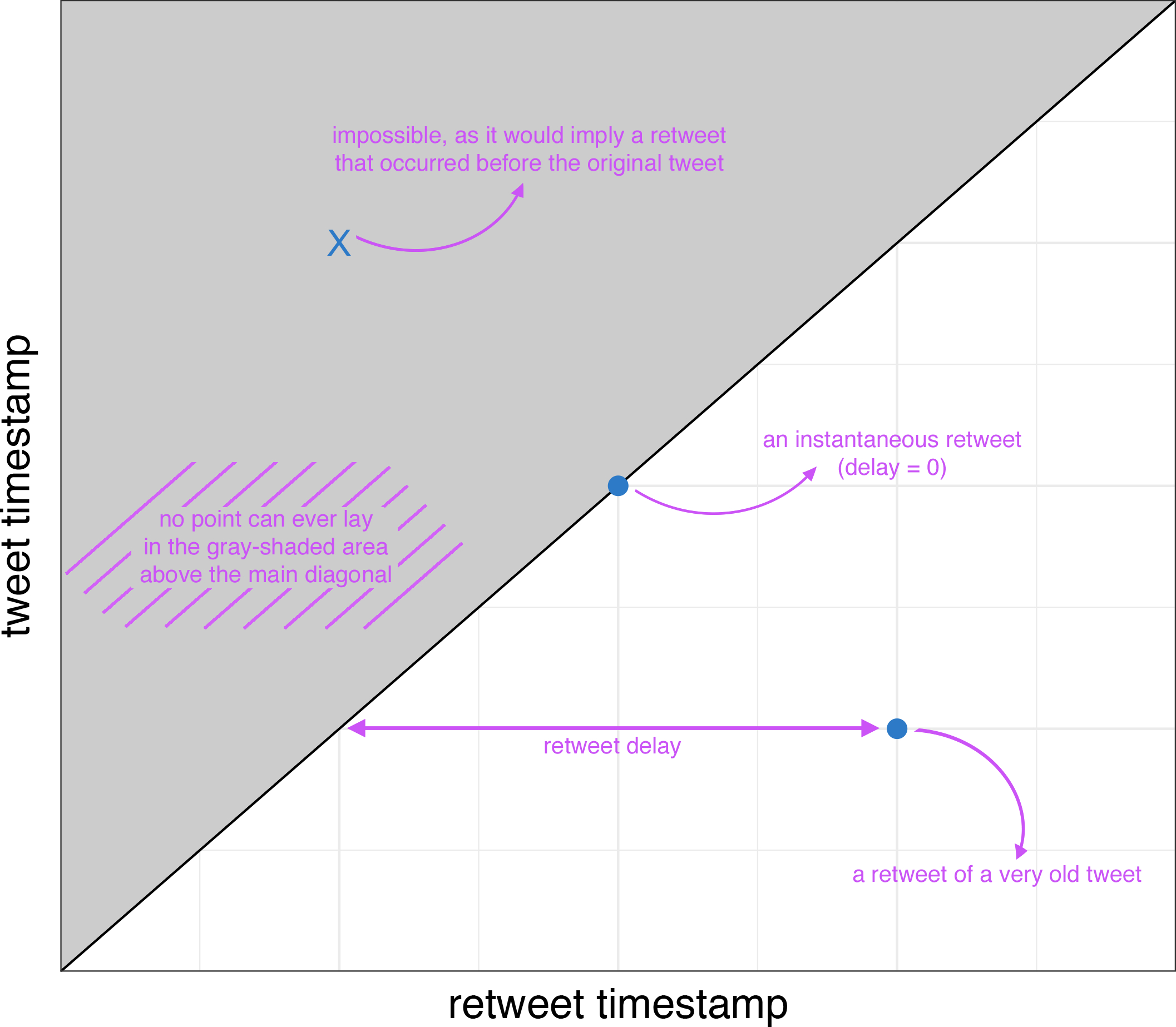}
        \caption{Semantics of RTT plots.\label{fig:RTT-explanation}}
	\end{subfigure}
	\hspace{0.075\textwidth}
	\begin{subfigure}[b]{0.25\textwidth}
        \centering
        \includegraphics[width=\textwidth]{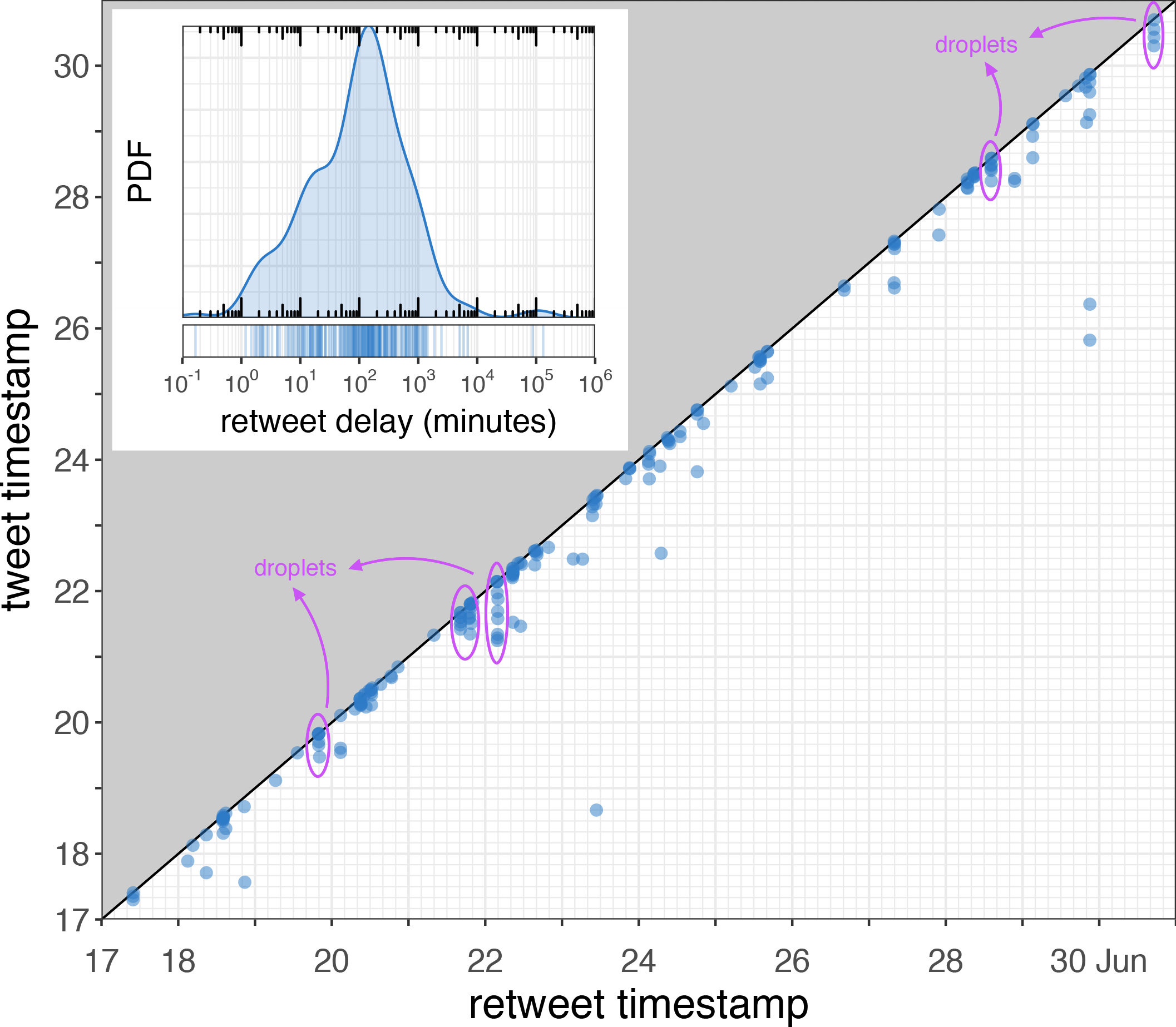}
        \caption{Droplet pattern.\label{fig:RTT-normal-droplet}}
	\end{subfigure}
	\hspace{0.075\textwidth}
	\begin{subfigure}[b]{0.25\textwidth}
        \centering
        \includegraphics[width=\textwidth]{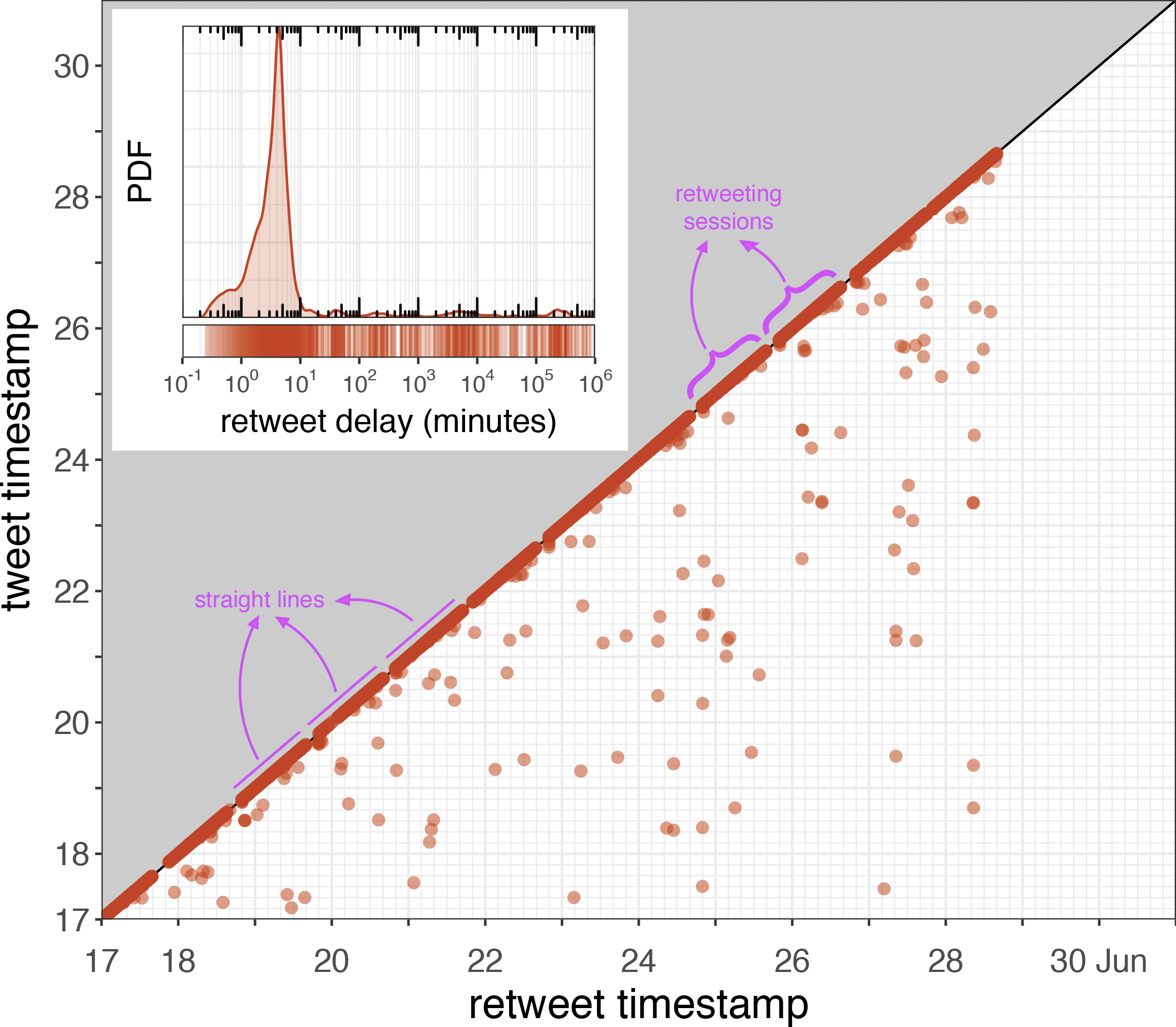}
        \caption{Straight line pattern.\label{fig:RTT-suspicious-straightline}}
	\end{subfigure}\\
	\begin{subfigure}[b]{0.25\textwidth}
        \centering
        \includegraphics[width=\textwidth]{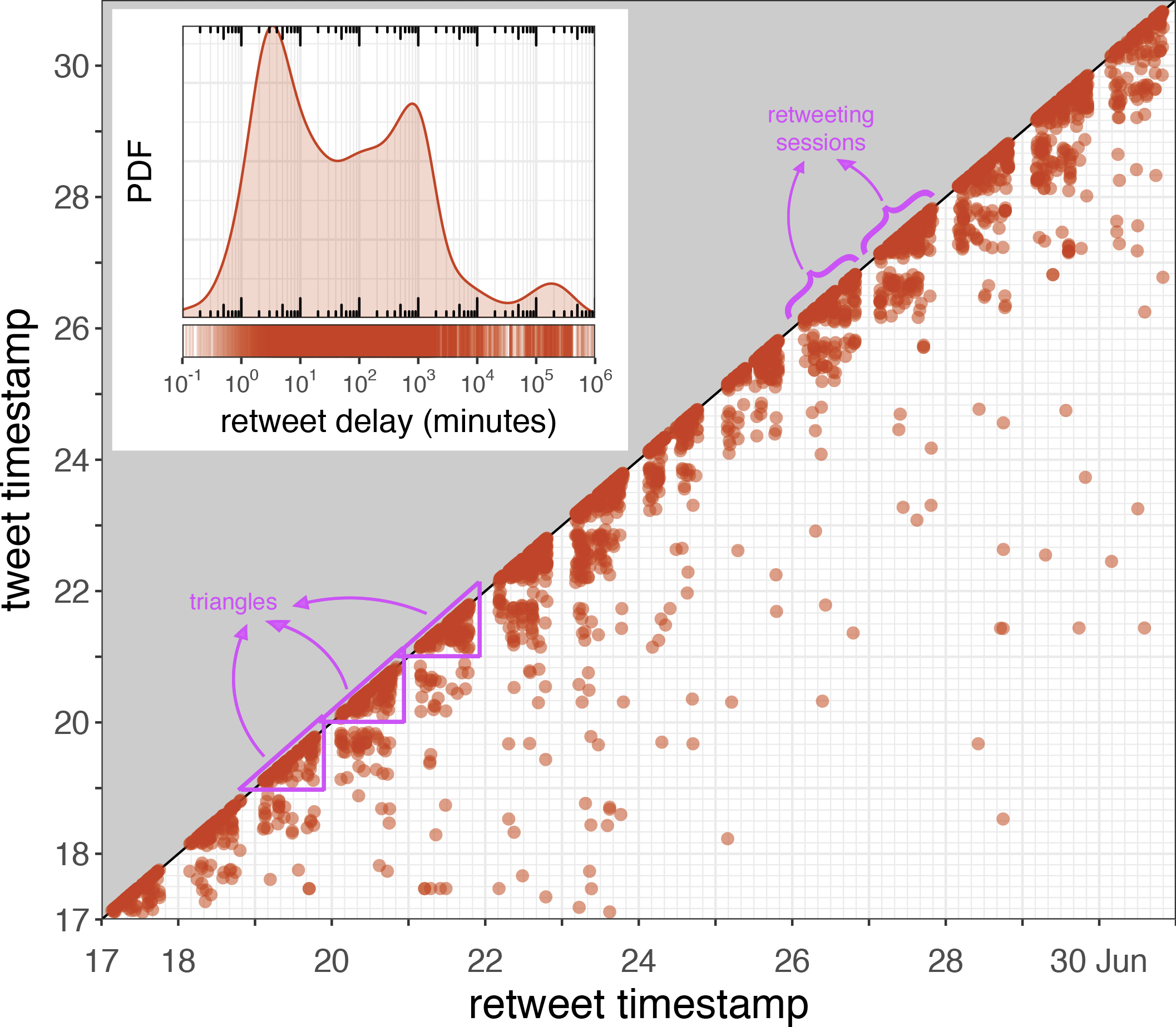}
        \caption{Triangular pattern.\label{fig:RTT-suspicious-triangle}}
	\end{subfigure}
	\hspace{0.075\textwidth}
	\begin{subfigure}[b]{0.25\textwidth}
        \centering
        \includegraphics[width=\textwidth]{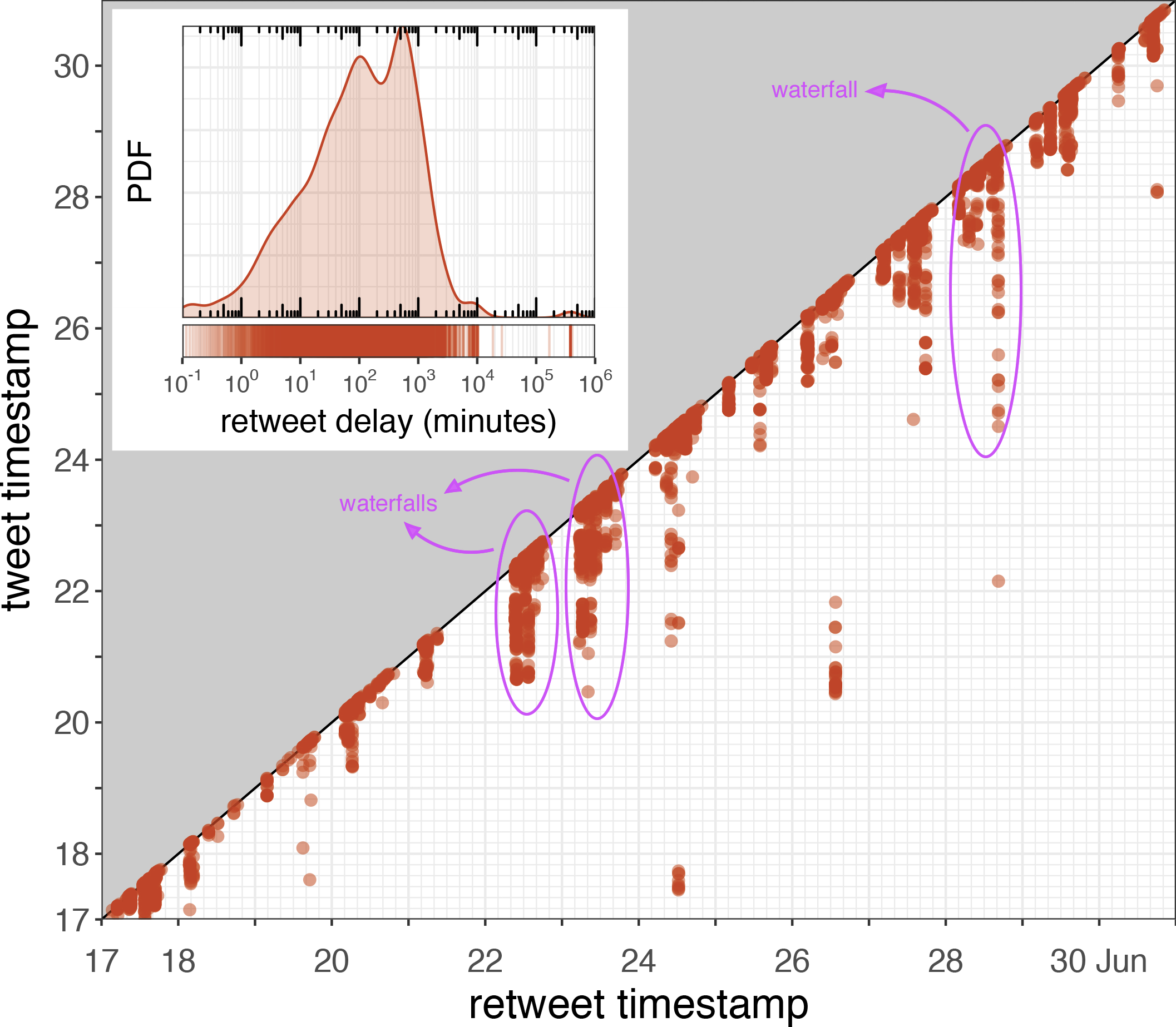}
        \caption{Waterfall pattern.\label{fig:RTT-suspicious-waterfall1}}
	\end{subfigure}
	\hspace{0.075\textwidth}
	\begin{subfigure}[b]{0.25\textwidth}
        \centering
        \includegraphics[width=\textwidth]{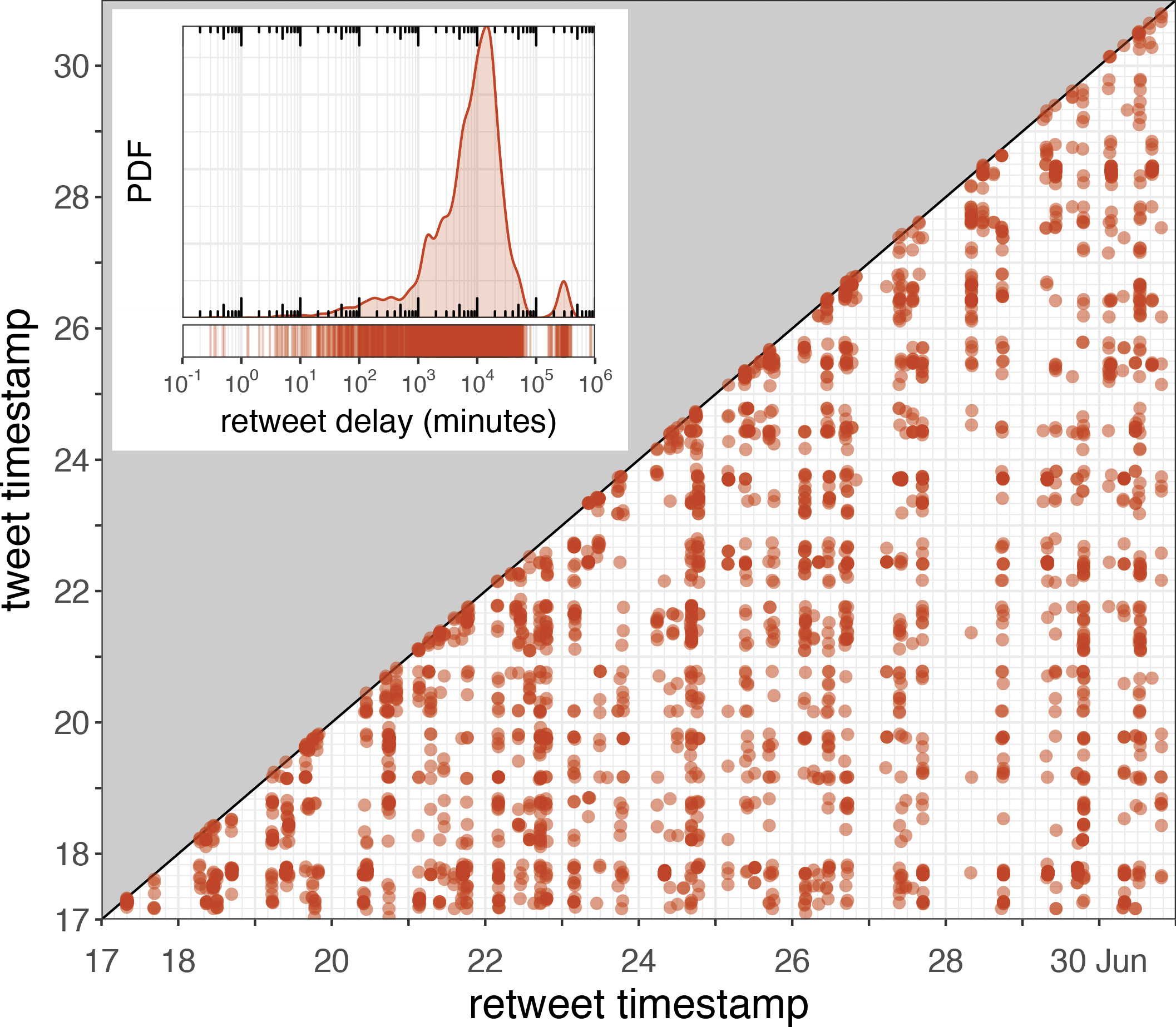}
        \caption{An extreme waterfall pattern.\label{fig:RTT-suspicious-waterfall2}}
	\end{subfigure}
	\caption{\textsc{\textmd{RTT}} plots depicting a normal tweeting behavior (blue colored) and many suspicious ones (red colored). The insets of \textsc{\textmd{RTT}} plots show the empirical probability density function (PDF) and a rug plot of retweet delays.\label{fig:RTT-plots}}
\end{figure*}

\section{Patterns of retweeting activity}
\label{sec:exploration}
Here, we investigate the temporal dynamics of retweeting activity of all the 63,762 accounts in our dataset. By also leveraging class labels of the annotated accounts, we aim at highlighting retweeting behaviors that are indicative of normal versus suspicious activity.

To ease the exploration of a user's retweeting activity, we propose a compact -- yet informative -- scatterplot visualization called \textsc{ReTweet-Tweet} (\textsc{RTT}). Given a user and the list of all his retweets, \textsc{RTT} plots the timestamp of each retweet ($x$ axis) against the timestamp of the corresponding original tweet ($y$ axis). By the definition of \textsc{RTT}, points laying in different areas of the plot imply different retweeting behaviors. Figure~\ref{fig:RTT-explanation} shows the semantics of this visualization, by means of a visual explanation of the most common behaviors caught by the \textsc{RTT} plot. In detail, no point can ever appear above the main diagonal of the plot, since that would break causality (i.e., it would correspond to a retweet that anticipates the original tweet, which is clearly impossible). Points laying near to the main diagonal represent retweets occurring rapidly after the publication time of the original tweet. Conversely, points that lay far from the diagonal imply a large temporal distance between a tweet and its retweet (i.e., retweets of very old tweets), which is an uncommon behavior.

\textbf{Normal behaviors -- droplet pattern.} Figure~\ref{fig:RTT-normal-droplet} shows the typical \textsc{RTT} plot of a ``normal'', legitimate user. As shown, the vast majority of points concentrates slightly below the main diagonal. This is the expected behavior on the past-paced Twitter OSN, since retweets typically occur with short delay (e.g., between 1 and 10 minutes) from the original tweets~\cite{gomez2014quantifying}. Occasionally, legitimate users retweet a sequence of tweets with increasing delays. In \textsc{RTT} plots, this retweeting behavior creates a few vertically-stacked points that we refer to as \textit{droplets}. We found this pattern to be frequent among legitimate users. In fact, the droplet pattern occurs each time a user retweets a Twitter feed (e.g., search results) or a user timeline. Since feeds and timelines are rendered -- both in the Web application and via APIs -- in reverse chronological order, retweeting a sequence of such tweets results in increasing retweet delays, just because tweets retweeted last are actually the oldest ones. Apart from these observations, no other clear pattern emerges in the \textsc{RTT} plot of Figure~\ref{fig:RTT-normal-droplet}. The possible presence of distinctive patterns in \textsc{RTT} plots is particularly relevant since it implies some form of regularity in a user's retweeting activity. In turn, striking regularities are typically caused by automated actions -- that is, they are representative of bot behaviors, as previous literature already highlighted~\cite{chavoshi2016debot,liu2017holoscope,cresci2017tdsc}.

\textbf{Suspicious behaviors -- straight line pattern.} The \textsc{RTT} plot in Figure~\ref{fig:RTT-suspicious-straightline} shows different retweeting behaviors. Almost all points in the plot are laying precisely above the main diagonal, meaning that the user almost always retweets in a matter of a few seconds from the original tweets. The user's activity is also clearly split into different sessions, separated by small gaps denoting inactivity. 
\textbf{Suspicious behaviors -- triangular pattern.} Figure~\ref{fig:RTT-suspicious-triangle} shows yet another suspicious pattern. This time the activity sessions are very regular, with always roughly the same length and the same inactivity time that separates subsequent sessions. Moreover, sessions seem to create a peculiar \textit{triangular} pattern below the diagonal of the \textsc{RTT} plot. This means that, within a given session, the user retweets past tweets only up to a fixed point in time that roughly corresponds to the starting time of the session.

\textbf{Suspicious behaviors -- waterfall pattern.} Finally, Figure~\ref{fig:RTT-suspicious-waterfall1} shows the behavior of a user whose retweets go way back in time. In figure, this is represented by points forming solid vertical lines. These lines are significantly longer than those representing droplets in Figure~\ref{fig:RTT-normal-droplet} and are probably caused by systematic retweeting of a Twitter feed or timeline in reverse chronological order. Since many vertical lines are present in the \textsc{RTT} plots showing this retweeting behavior, we call this a \textit{waterfall} pattern. Apart from the vertical lines, mild signs of retweeting sessions are also visible in the plot. Although the presence of the \textit{waterfall} pattern shown in Figure~\ref{fig:RTT-suspicious-waterfall1} is already indicative of automated behaviors, for some users this behavior is truly extreme, as shown in Figure~\ref{fig:RTT-suspicious-waterfall2}.

As highlighted by the previous analyses, the \textsc{RTT} plot is useful for studying the retweeting behavior of a given user. It is a valuable tool \textit{per se}, that can empower human analysts for getting insights into the behaviors of Twitter accounts. Moving forward, in the next section we build on the results of these analyses by designing a fully automatic technique for spotting malicious retweeting bots. 
\makeatletter{}\section{Introducing Retweet-Buster}
\label{sec:detection}
As highlighted in the previous sections, unsupervised group-based approaches currently represent the most promising research direction for social bot detection~\cite{Cresci2017,yang2019arming}. Thus, in this section we propose an unsupervised technique that leverages the temporal distribution of the retweets of large groups of users, in order to detect malicious retweeting bots. Since we aim to design an unsupervised technique, we can not simply learn a detector for spotting the suspicious patterns identified in Section~\ref{sec:exploration}. Instead, our goal is to develop a technique that is capable of spotting such suspicious patterns, \textit{as well as possible other patterns} that might exist in a dataset describing temporal retweeting activities. In other words, the technique should be able to automatically identify meaningful patterns from the data, rather than be able to recognize only those patterns that we manually labeled as suspicious.

\begin{figure}[t]
    \centering
    \includegraphics[width=0.85\columnwidth]{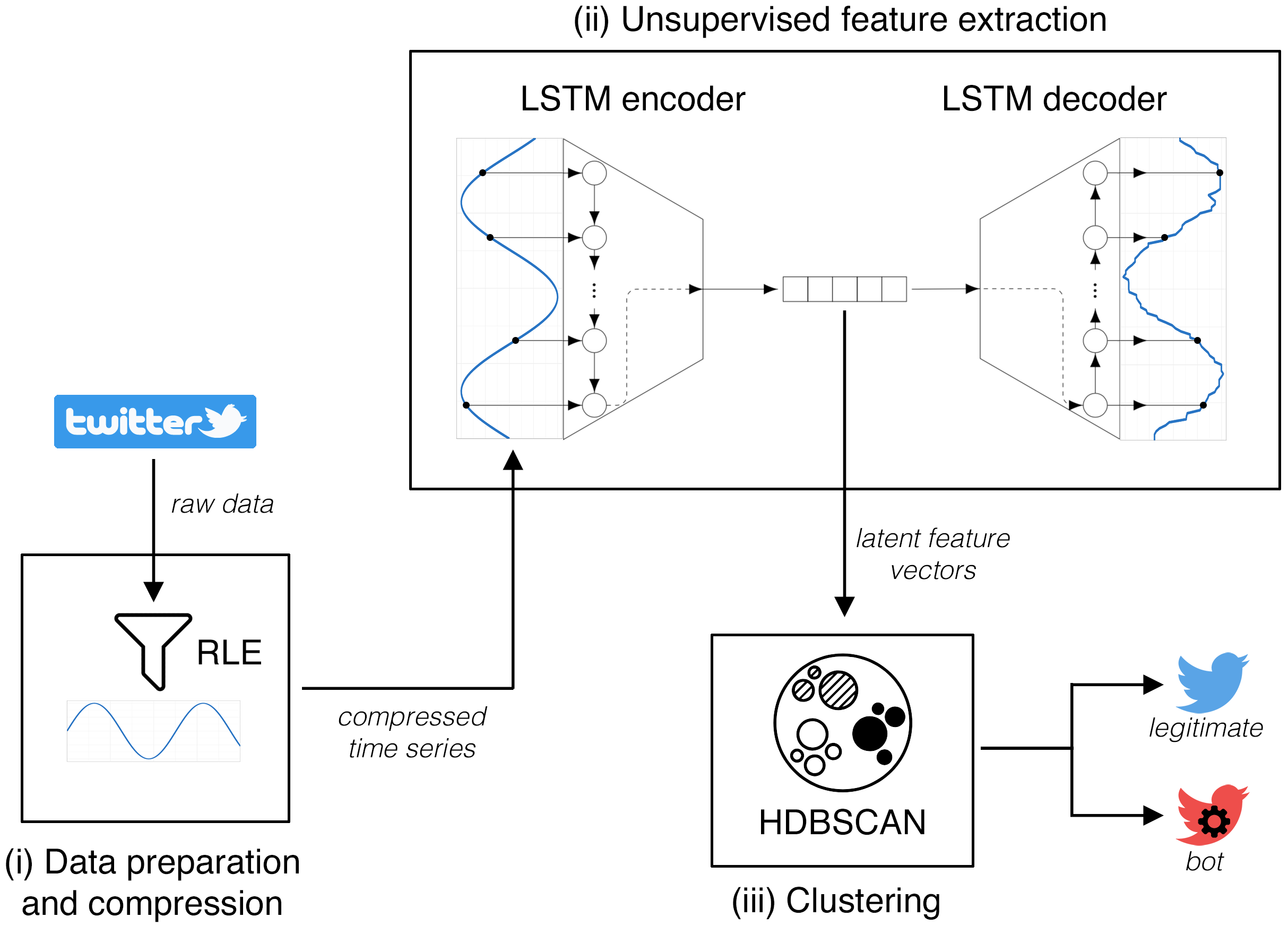}
    \caption{Main logical components of \textsc{\textmd{RTbust}}.\label{fig:RTbust}}
\end{figure}

Thus, building on these considerations, we propose a retweeting bot detection technique, called \textsc{Retweet-Buster} (\textsc{RTbust}), based on automatic unsupervised feature extraction. Then, in order to implement a group-analysis technique, such automatically-learned features are passed to a density-based clustering algorithm. The rationale for clustering stems from previous research in human and bot behaviors, in that humans have been proven to exhibit much more behavioral heterogeneity than automated accounts~\cite{chavoshi2016debot,dsaa2017}. As a consequence, we expect that the heterogeneous humans will not be sufficiently ``dense'' to be clustered. In fact, they will act pretty much like a background noise in the features space. Conversely, groups of coordinated and synchronized bots will be organized in much denser groups in the features space, thus resulting in nice density-based clusters. In detail, our proposed \textsc{RTbust} technique is organized in 3 main steps, as also shown in Figure~\ref{fig:RTbust}: (i) time series data preparation and compression; (ii) unsupervised feature extraction; and (iii) user clustering.

\textbf{Data preparation and compression.} Let $t_{REF}$ be a reference timestamp set at the start of the analysis time window. In our case $t_{REF}$ corresponds to 17 June, 2018 at 00:00:00. Then, $t(x)$ denotes the publication timestamp of tweet ID $= x$. The raw data exploited to carry out bot detection with \textsc{RTbust} is the same used in the \textsc{RTT} plots of Section~\ref{sec:exploration}, that is tweet and retweet timestamps. In \textsc{RTbust} such timestamps are organized as a retweet time series $R_i = \{r_{i,0}, r_{i,1}, \ldots, r_{i,n}\}$ for each user $u_i$, where,
\begin{equation*}
\label{eq:ts-obs}
r_{i,j} =
\begin{cases} 
	|t(x) - t_{REF}|	& \text{if } u_i \text{ retweeted tweet ID} = x \text{ at time } t_j \\
	0	& \text{if } u_i \text{ did not retweet at time } t_j \\
\end{cases}
\end{equation*}
The temporal granularity of our time series is that of seconds, which is the same granularity as Twitter timestamps. That is, we have one $r_{i,j}$ observation per second per user. Hence, our time series have a very fine temporal grain. However, overall they are very sparse because users retweet, on average, only once in a few minutes.

\begin{figure}[t]
    \centering
    \resizebox{0.75\columnwidth}{!}{        \makeatletter{}\begin{tikzpicture}[cell/.style = {rectangle, draw = black}, nodes in empty cells]
    \matrix[
  matrix of nodes,
  row sep = 0.35em,
  column sep = -\pgflinewidth,
  nodes={anchor = center, text height = 2ex, text depth = 0.25ex, cell},
  column 1/.style = {nodes={minimum width = 2em, draw = none}},
  column 2/.style = {nodes={minimum width = 2em}},
  column 3/.style = {nodes={minimum width = 2em}},
  column 4/.style = {nodes={minimum width = 2em}},
  column 5/.style = {nodes={minimum width = 2em}},
  column 6/.style = {nodes={minimum width = 2em}},
  column 7/.style = {nodes={minimum width = 2em}},
  column 8/.style = {nodes={minimum width = 2em}},
  column 9/.style = {nodes={minimum width = 2em}},
  column 10/.style = {nodes={minimum width = 2em}},
  column 11/.style = {nodes={minimum width = 2em}},
  row 1/.style={nodes={draw = none}},
  ampersand replacement=\&,
  ] (top) {
    \& $r_{i,0}$ \& $r_{i,1}$ \& $\cdots$ \&\&\&\&\&\& $\cdots$ \& $r_{i,n}$ \\
    $R_i$ = \& 3 \& |[text = orange]|0 \& |[text = orange]|0 \& |[text = orange]|0 \& 4 \& |[text = orange]|0 \&6 \& 9 \& $\cdots$ \& 20 \\
  };

    \matrix[
  matrix of nodes,
  below = 2em of top,
  row sep = 0.35em,
  column sep = -\pgflinewidth,
  nodes={anchor = center, text height = 2ex, text depth = 0.25ex, cell},
  column 1/.style = {nodes={minimum width = 2em, draw = none}},
  column 2/.style = {nodes={minimum width = 2em}},
  column 3/.style = {nodes={minimum width = 2em}},
  column 4/.style = {nodes={minimum width = 2em}},
  column 5/.style = {nodes={minimum width = 2em}},
  column 6/.style = {nodes={minimum width = 2em}},
  column 7/.style = {nodes={minimum width = 2em}},
  column 8/.style = {nodes={minimum width = 2em}},
  column 9/.style = {nodes={minimum width = 2em}},
  ampersand replacement=\&,
  ] (bottom) {
    \& 3 \&|[text = orange]|-3 \& 4 \& |[text = orange]|-1 \& 6 \& 9 \& $\cdots$ \& 20 \\
  };
  
    \draw [dashed] (top-2-2.south west) -- (bottom-1-2.north west);
  \draw [dashed] (top-2-3.south west) -- (bottom-1-3.north west);
  \draw [dashed] (top-2-6.south west) -- (bottom-1-4.north west);
  \draw [dashed] (top-2-7.south west) -- (bottom-1-5.north west);
  \draw [dashed] (top-2-8.south west) -- (bottom-1-6.north west);
  \draw [dashed] (top-2-9.south west) -- (bottom-1-7.north west);
  \draw [dashed] (top-2-10.south west) -- (bottom-1-8.north west);
  \draw [dashed] (top-2-11.south west) -- (bottom-1-9.north west);
  \draw [dashed] (top-2-11.south east) -- (bottom-1-9.north east);
\end{tikzpicture} 
    }
    \caption{Excerpt of a retweet time series compressed with RLE. Observations affected by RLE are orange-colored.\label{fig:RLE}}
\end{figure}
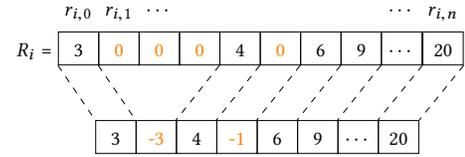

At this step, we want to maximize the informativeness of our time series, while minimizing the amount of data to store and process. To beat this trade off, we employ a modified version of the run-length encoding (RLE) compression scheme. RLE is a simple and widely-used lossless sequence compression scheme that encodes consecutive equal data values with one single \textit{<value, length>} tuple. In our case, we use RLE to reduce the sparsity in our time series by compressing the many consecutive zeros that represent no retweeting activity. Since we only need to compress zeros, we do not need to use the typical RLE \textit{<value, length>} tuples. In fact, just the \textit{length} will suffice. We have however to make a distinction between RLE lengths and the $|t(x) - t_{REF}|$ observations that represent retweets. Since by definition, $r_{i,j}$ observations that represent retweets are always positive, one straightforward way to make such distinction is to substitute consecutive zeros with the opposite of RLE lengths. Figure~\ref{fig:RLE} graphically summarizes this data preparation and compression step. 
As a result of this step, we have one single compact time series for each user. Furthermore, such time series efficiently encodes multiple information, such as inactivity periods as well as tweets and retweets timestamps.

\textbf{Unsupervised feature extraction.} So far, we obtained a compact representation of the temporal retweeting behaviors of our users. Now, we need to turn this representation into a limited number of highly-informative features. Specifically, we have 3 goals for this step: (i) obtain features for a subsequent machine learning task (i.e., density-based clustering) in an unsupervised way; (ii) obtain fixed-length feature vectors, whereas user time series have variable length; (iii) maximize the amount of information in our features, while minimizing the number of features.

Basically, all our goals can be achieved with a well thought out application of dimensionality reduction techniques to our run-length-encoded time series. Specifically, we take a feature projection approach, where we aim to transform our compressed time series into lower-dimensional feature vectors. Feature projection can be achieved via a large number of different techniques. Here, we propose a solution based on \textit{variational autoencoders}, a particular type of deep neural networks, because of their favorable characteristics~\cite{kingma2013auto}. In Section~\ref{sec:results} we also show comparisons with other dimensionality reduction and feature projection techniques (e.g., PCA, TICA).

Variational autoencoders (VAEs) are an unsupervised learning technique that leverages neural networks for learning a probabilistic representation (i.e., learning features) of the input. As shown in Figure~\ref{fig:RTbust}, a VAE is composed of 2 main modules: an encoder network and a decoder network. The encoder takes the original input and learns a compressed knowledge representation. Dually, the decoder takes the compressed representation and tries to reconstruct the original input at its best. During the (unsupervised) training phase of the network, the encoder learns to create a meaningful and informative compressed representation of the input, so that the decoder can accurately reconstruct it. Notably, if the input dimensions are largely independent of one another, the compression performed by the encoder results in a loss of valuable information and the subsequent reconstruction becomes a very difficult task. However, if some sort of structure (i.e., patterns) exists in the input data, such as the peculiar patterns that we uncovered in Section~\ref{sec:exploration}, that structure can be learned by the encoder and consequently leveraged when converting the input into its compressed representation. In other words, this encoding learns latent features of the input data, which is precisely what we require from this unsupervised feature extraction step. Because deep neural networks are capable of learning nonlinear relationships, this encoding can be thought of as a more powerful generalization of PCA~\cite{meng2017relational}. In our work, the decoder is only used for training the VAE, since we are not really interested in reconstructing our time series. Instead, once trained, the encoder becomes our unsupervised feature extractor. The dimension of the latent feature vector generated by the encoder is a parameter of the VAE, with which we extensively experiment in the next section. The lower the dimension, the more compressed is the representation of the input. However, a low-dimensional representation mitigates possible issues during the subsequent clustering step, caused by the curse of dimensionality~\cite{domingos2012few}.

Regarding the deep learning architecture used for implementing the VAE, we relied on a long short-term memory (LSTM) network. LSTMs are well-known and widely used recurrent neural networks. Because of their memory states, they are well-suited for performing tasks on time series, which are often characterized by temporal correlations and by lags of unknown duration between relevant events~\cite{guo2016robust}. Specifically, LSTMs have been proven very accurate at extracting patterns in input space, where input data spans over long sequences. Furthermore, they are also suitable for dealing with sequences of variable length, such as our time series, while still being able to produce fixed-lengths data representations.

\textbf{User clustering.} Now that our users are represented by their latent feature vectors computed by the LSTM encoder, we can apply density-based clustering in order to check for common retweeting behaviors. If large clusters of users are found, then we might have detected a coordinated and synchronized group of accounts, possibly constituting a retweeting botnet. We base our clustering step on a recent efficient algorithm that combines density- and hierarchical-based clustering: HDBSCAN~\cite{campello2013density}. Among the advantages of this algorithm is its effectiveness in finding clusters with variable degrees of density, a much desirable feature when dealing with noisy real-world data. In addition, HDBSCAN also proved about twice as fast as its predecessor DBSCAN. Regarding algorithm parameters, HDBSCAN removed the need to specify a global density threshold (the $\epsilon$ parameter in DBSCAN) by employing an optimization strategy for finding the best cluster stability~\cite{campello2013density}. The only mandatory parameter to set is the minimum cardinality of the clusters found by HDBSCAN. We experiment with this parameter in the next section. Concluding, after running our density-based clusering step, we label as bots all those accounts that end up clustered. Conversely, we label as legitimate all those account that are treated as noise (i.e., that are not clustered) by HDBSCAN.
\makeatletter{}\section{Experiments and Results}
\label{sec:results}
Here we present results obtained while searching for the best set of parameters of \textsc{RTbust}, as well as overall bot detection results.

\textbf{Evaluation methodology.} 
In all the experiments described in this section we analyze all 63,762 accounts of our dataset. Then, we evaluate the effectiveness of different techniques, possibly executed with different configurations of parameters, in correctly classifying the $\simeq 1,000$ annotated accounts. Thus, the bot detection task is framed as a binary classification task, with the 2 classes being: \textit{bot} (positive class) and \textit{human} (negative class). For presenting evaluation results we rely on 5 standard, well-known metrics used for evaluating binary machine learning classifiers, specifically: precision, recall, accuracy, F1-Score ($F1$), and Matthews correlation coefficient ($MCC$)~\cite{powers2011}.

\textbf{Comparisons.}
While evaluating the bot detection results of \textsc{RTbust}, we also perform extensive comparisons with other baselines and state-of-the-art techniques for social bot detection.

In Section~\ref{sec:detection} we grounded the unsupervised feature extraction step of \textsc{RTbust} on a variational autoencoder. However, a number of other techniques could be used to achieve the same goal. Thus, for the sake of experimentation we implemented 2 more versions of \textsc{RTbust} that are based respectively on principal component analysis (PCA) and time independent component analysis (TICA) for feature extraction~\cite{hernandez2018variational}. PCA is a well-known linear statistical technique for dimensionality reduction. By choosing an appropriate number of principal components found by PCA, it is possible to greatly reduce the dimensionality of a problem, while minimizing the loss of information. Similarly, TICA is another dimensionality reduction technique that is particularly suitable for dealing with time series. While PCA finds high-variance linear combinations of the input dimensions, TICA aims to find high-autocorrelation linear combinations. Thus, when using PCA and TICA for feature extraction, the features resulting from PCA are likely to convey different information with respect to those obtained from TICA.

\begin{table}[t]
	\small
	\centering
	\begin{tabular}{llp{0.6\columnwidth}}
		\toprule
		& \textbf{feature} & \textbf{description} \\
		\midrule
            1 & RT users entropy & Shannon entropy of the distribution of retweeted users \\
            2 & RT days entropy & Shannon entropy of the distribution of the publication days of retweeted tweets \\
            3 & RT rate & number of retweets per time unit \\
            4 & daily mean RTs & daily mean of the number of retweets \\
            5 & RT days & number of distinct days during which the user retweeted at least once \\
            6 & minimum IRT & minimum of the inter-retweet times \\
            7 & mean IRT & mean of the inter-retweet times \\
            8 & stdev IRT & standard deviation of the inter-retweet times \\
            9 & minimum RT delay & minimum of the retweet delays \\
            10 & mean RT delay & mean of the retweet delays \\
            11 & stdev RT delay & standard deviation of the retweet delays \\
            12 & RT sessions & number of detected retweeting sessions \\
		\bottomrule
	\end{tabular}
	\caption{List of handcrafted features computed for each user.\label{tab:handcrafted-feat}}
\end{table}

As typically done for many machine learning tasks, we also experiment with a small set of handcrafted features. These features take into account characteristics of the retweet time series of the users (e.g., inter-retweet times, retweeting delays and sessions), as well as other characteristics not available from retweet time series, but that might still contribute to bot detection (e.g., the distribution of retweeted accounts). Table~\ref{tab:handcrafted-feat} lists the 12 handcrafted features. Notably, some of them -- such as inter-retweet times -- have been largely used in previous retweeter bot detection systems~\cite{vo2017revealing,gupta2019malreg}. In our subsequent experiments, these handcrafted features are tested in \textsc{RTbust} in place of those automatically extracted by PCA, TICA, and the VAE.

Regarding comparisons with other state-of-the-art bot detection techniques, we experiment with the supervised version of the \textit{Social fingerprinting} technique proposed in~\cite{cresci2016dna,cresci2017tdsc} and with the unsupervised \textit{HoloScope} technique proposed in~\cite{liu2017holoscope}. Both techniques are designed to detect synchronized and coordinated groups of bots, and have been briefly described in Section~\ref{sec:relwork}. In addition, we also experiment with the \textit{Botometer} system~\cite{davis2016}. \textit{Botometer} is a publicly-available service\footnote{\scriptsize{\url{https://botometer.iuni.iu.edu}}} to evaluate the similarity of a Twitter account with the known characteristics of social bots. It leverages an off-the-shelf supervised machine learning classifier that exploits more than 1,000 features of the accounts under investigation. Similarly to the majority of existing systems, \textit{Botometer} performs account-by-account analyses, rather than group analyses.

Finally, as a simple comparison baseline we also try classifying as bots all those accounts whose retweet rate (i.e., retweets per unit of time) is higher than a fixed threshold. We set this threshold as the third quartile of the distribution of retweet rates for our 63,762 accounts. This comparison helps understand if simply looking at the number of retweets is enough for gaining insights into the nature (bot/legitimate) of an account.

\begin{figure}[t]
    \centering
    \includegraphics[width=0.9\columnwidth]{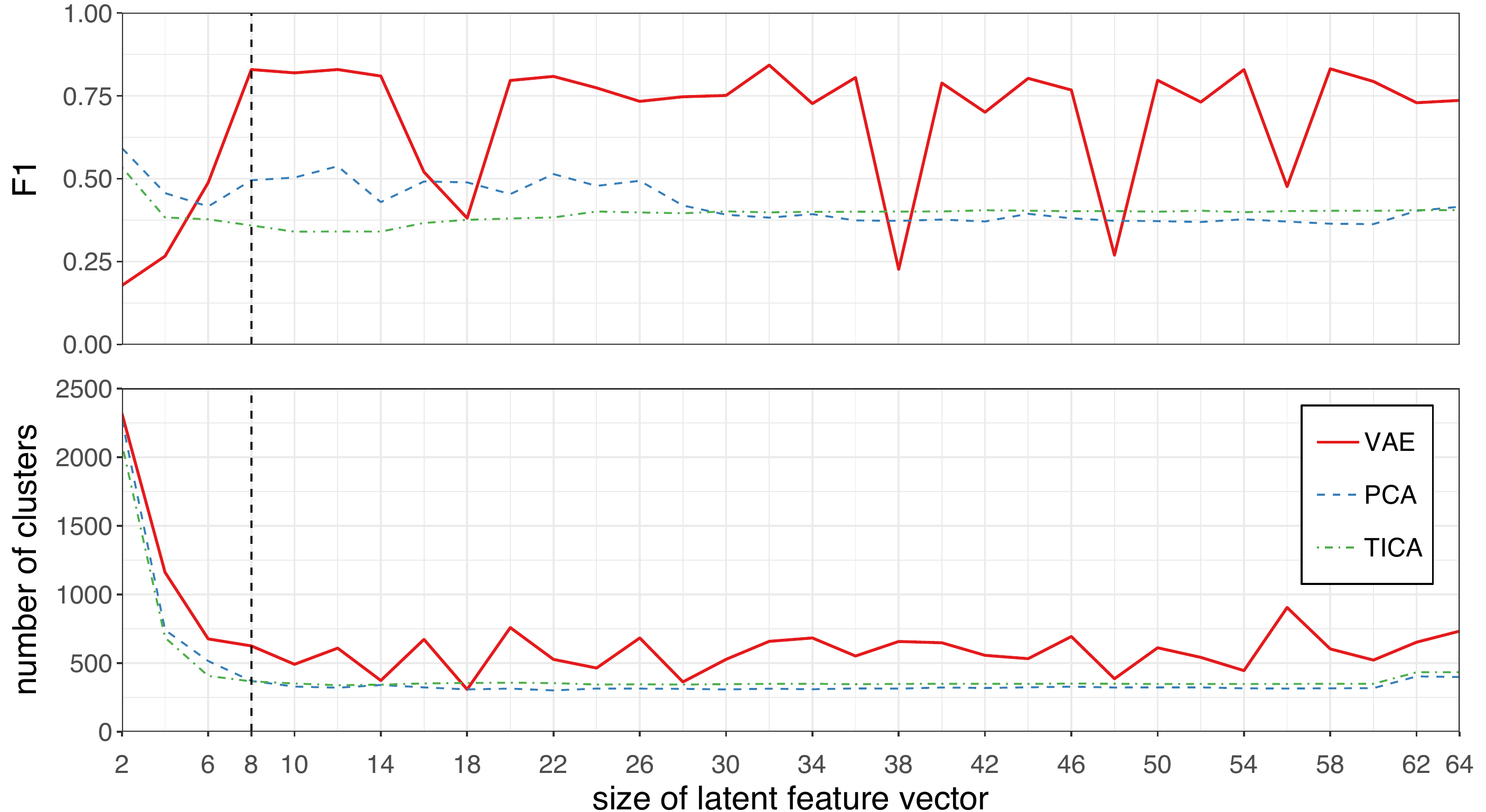}
    \caption{Dependence of clustering stability and bot detection performances on the number of features extracted by PCA, TICA and the VAE. Best results are obtained with the VAE, starting from 8 features (dashed vertical line).\label{fig:MC_simulation}}
\end{figure}

\textbf{Parameters configuration.}
The most important parameter of \textsc{RTbust} is the dimension of the latent feature vectors generated by the VAE. It affects both the quality of the subsequent clustering step, which in turn affects bot detection performances, as well as the time/computation needed to complete it. The same also applies to the number of projected dimensions obtained with PCA or TICA. Thus, we designed an experiment to evaluate the performances of \textsc{RTbust}, implemented with VAE, PCA, and TICA for feature extraction, in relation to the number of considered latent features.

Figure~\ref{fig:MC_simulation} shows results for this experiment, in terms of the overall $F1$ obtained for the classification of bots and humans, and of the total number of clusters found by HDBSCAN. Regarding the number of clusters, all 3 implementations show the same qualitative behavior. When considering a very limited number of features, HDBSCAN finds a large number of clusters. Increasing the number of features results in fewer clusters, up to a point (size of latent feature vector $= 10$) where the number of clusters stops decreasing. Regarding bot detection performance, Figure~\ref{fig:MC_simulation} shows a very different behavior between the VAE implementation of \textsc{RTbust}, with respect to the PCA and TICA ones. Features extracted by PCA and TICA seem to provide significantly less information for bot detection with respect to those extracted by the VAE, as demonstrated by a lower $F1$. Moreover, increasing the number of considered PCA and TICA features seems of no help, since the $F1$ remains almost the same for each latent feature vector size. Instead, the VAE implementation shows a behavior that is consistent with the results obtained for the number of clusters. The $F1$ rapidly increases when considering a larger number of features, up to a point (size of latent feature vector $= 8$) where it reaches and maintains its maximum (apart from a few fluctuations). Overall, these results demonstrate that the VAE is much more powerful than PCA and TICA for the unsupervised feature extraction from our retweet time series. Furthermore, as little as 8 VAE features seem to be enough for stabilizing the clustering and achieving good bot detection results.

We repeated this experiment by also varying the minimum cluster cardinality (i.e., the only HDBSCAN parameter). We omit detailed results due to space limitations, however we report finding the best results with clusters larger than 10 accounts, a threshold that is both intuitive (i.e., it is not very meaningful to look for minuscule botnets) and operationally effective.

\makeatletter{}\begin{table*}[t]
	\small
	\centering
	\begin{tabular}{llrrrrrr}
		\toprule
		&&& \multicolumn{5}{c}{\textbf{evaluation metrics}} \\
		\cmidrule{4-8}
		\textbf{technique} & \textbf{type} & \textbf{features} & \textit{precision} & \textit{recall} & \textit{accuracy} & $F1$ & $MCC$ \\
		\midrule
		\multicolumn{8}{l}{\textit{baseline}} \\ [0.8ex]
			retweet rate                            & --		        & 1 & 0.3534	& 0.3585	& 0.3440	& 0.3559	& $-$0.3124 \\
		\midrule
		\multicolumn{8}{l}{\textit{comparisons}} \\ [0.8ex]
			Botometer~\cite{davis2016}                  & supervised	& $> 1,000$ & 0.6951	& 0.3098	& 0.5830	& 0.4286	& 0.2051 \\
			HoloScope~\cite{liu2017holoscope}           & unsupervised	& -- & 0.2857	& 0.0049	& 0.4908	& 0.0096	& $-$0.0410 \\
			Social fingerprinting~\cite{cresci2016dna,cresci2017tdsc} & supervised	& -- & 0.6562	& 0.8978	& 0.7114	& 0.7582	& 0.4536 \\
		\midrule
		\multicolumn{8}{l}{\textit{our contributions}} \\ [0.8ex]
			\textsc{RTbust} (handcrafted features)		& unsupervised		& 12 & 0.5284	& 0.7707	& 0.5364	& 0.6270	& 0.0767 \\
			\textsc{RTbust} (PCA)		            & unsupervised		& 8 & 0.5111	& \textbf{0.9512}	& 0.5154	& 0.6649	& 0.0446 \\
			\textsc{RTbust} (TICA)		            & unsupervised		& 8 & 0.5228	& \textbf{0.9512}	& 0.5364	& 0.6747	& 0.1168 \\
            \textsc{RTbust} (VAE)		            & unsupervised		& 8 & \textbf{0.9304}	& 0.8146	& \textbf{0.8755}	& \textbf{0.8687}	& \textbf{0.7572} \\
		\bottomrule
	\end{tabular}
	\caption{Retweeter bot detection results of \textsc{\textmd{RTbust}} and comparison with a baseline and other state-of-the-art techniques. Best results in each evaluation metric are shown in bold.\label{tab:results}}
\end{table*} 

\textbf{Quantitative evaluation of bot detection.} Next, we present detailed bot detection results for all the considered techniques. For the VAE, PCA and TICA implementations of \textsc{RTbust}, we use only 8 features, leveraging results of our previous experiment.

Table~\ref{tab:results} shows retweeter bot detection results. The best detection performances ($F1 = 0.87$) are achieved by the proposed \textsc{RTbust} technique using the VAE for unsupervised feature extraction. In fact, it beats all other competitors in each evaluation metric, with the only exception of the \textit{recall} metric. All other implementations of \textsc{RTbust} achieve worse results, with $F1 \leq 0.67$. As anticipated, this is a strong point in favor of the VAE for extracting informative features from our retweet time series. The second best overall results are obtained by the \textit{Social fingerprinting}, achieving an encouraging $F1 = 0.76$. Instead, the other state-of-the-art techniques obtain much worse results, with \textit{Botometer}'s $F1 = 0.43$ and \textit{HoloScope}'s $F1 = 0.01$. Interestingly, the majority of evaluated techniques achieve their worst results in the \textit{precision} metric, meaning that many legitimate accounts are misclassified as bots. This results is in contrast with previous results in bot detection~\cite{Cresci2017}. However, previous works mainly experimented with supervised techniques, while here we mainly explore unsupervised ones. Thus, combined results of our study and previous ones might suggest that supervised approaches to social bot detection are more prone to type II errors (false negatives), while unsupervised approaches are more prone to type I errors (false positives).

\begin{figure}[t]
    \centering    \includegraphics[width=0.9\columnwidth]{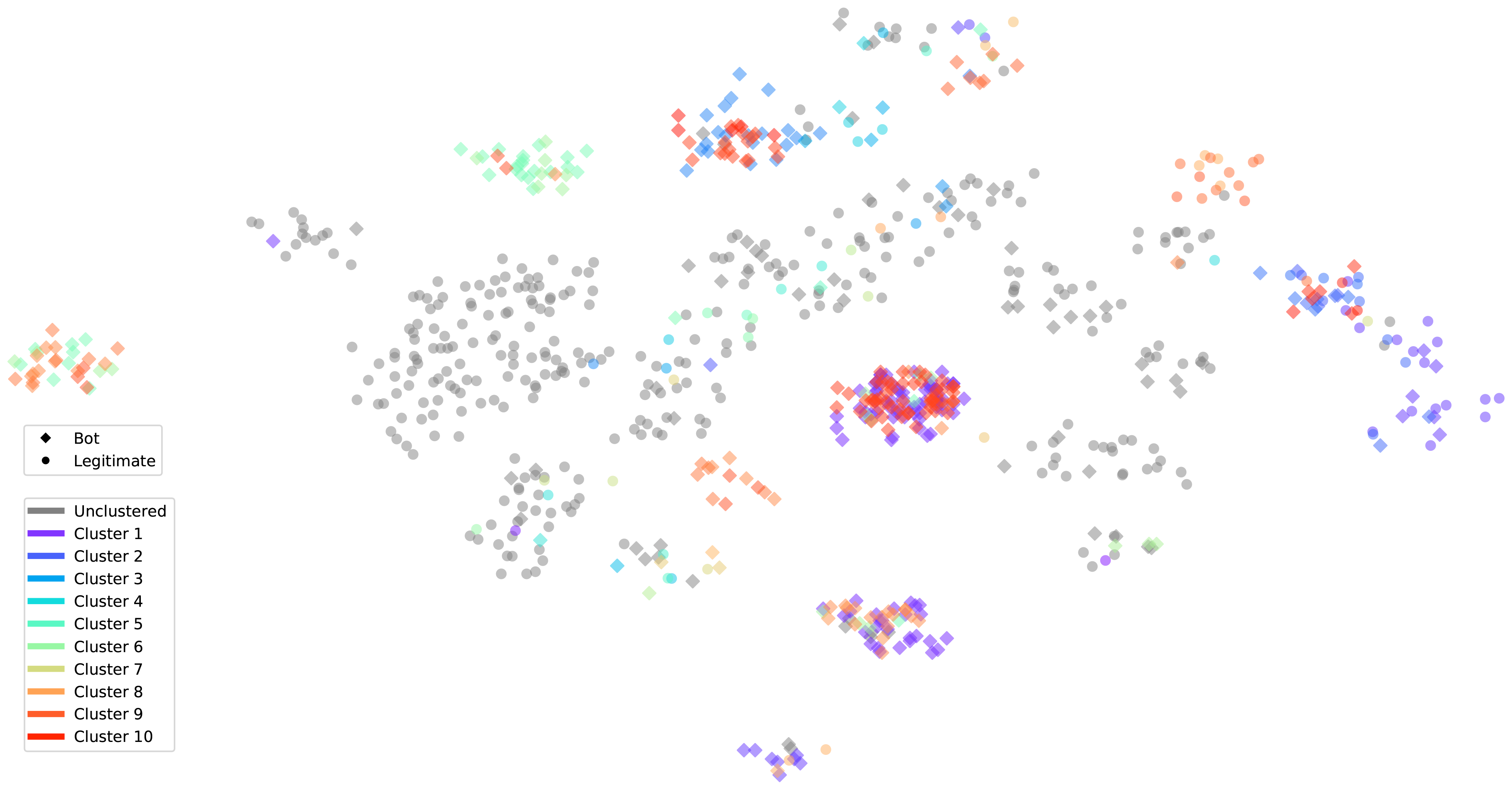}	\caption{Visual exploration of \textsc{\textmd{RTbust}} (VAE) bot detection results, and comparison with t-SNE.\label{fig:tSNE-detection}}
\end{figure}

  In Figure~\ref{fig:tSNE-detection}, we use t-SNE~\cite{maaten2008visualizing} for plotting our annotated accounts in a bi-dimensional space, where each account is colored according to the \textsc{RTbust} (VAE) cluster it belongs to, and 2 different symbols represent bots and legitimate accounts according to our ground truth. Since in \textsc{RTbust} each clustered account is labeled as bot while unclustered ones are labeled as humans, by comparing colors and symbols of the accounts in figure, it is possible to visually assess the quality of bot classification. Furthermore, Figure~\ref{fig:tSNE-detection} also allows a comparison between our clustering and that resulting from t-SNE. Regarding bot detection, we can see few clustered legitimate accounts. Similarly, only a minority of bots are grey-colored (i.e., unclustered). Concerning the comparison between t-SNE and \textsc{RTbust}, the 2 algorithms produce rather different clusters. For instance, multiple \textsc{RTbust} clusters collapse in a single t-SNE cluster and vice versa. However, the overall distinction between bots and humans seem to hold for both \textsc{RTbust} and t-SNE. The majority of legitimate accounts are both grey-colored and spread across the central region of Figure~\ref{fig:tSNE-detection}. In other words, they do not belong to any cluster neither for t-SNE nor for \textsc{RTbust}. Dually, the majority of bots appear as colored and positioned near other bots. Thus, both t-SNE and \textsc{RTbust} seem to recognize bot similarities although organizing the bots differently, despite the very different inner functioning of the 2 techniques. Given the usefulness and widespread application of t-SNE, these similarities support results of \textsc{RTbust} (VAE).

\begin{figure}[t]
	\begin{subfigure}[b]{0.24\textwidth}\centering\includegraphics[width=\textwidth]{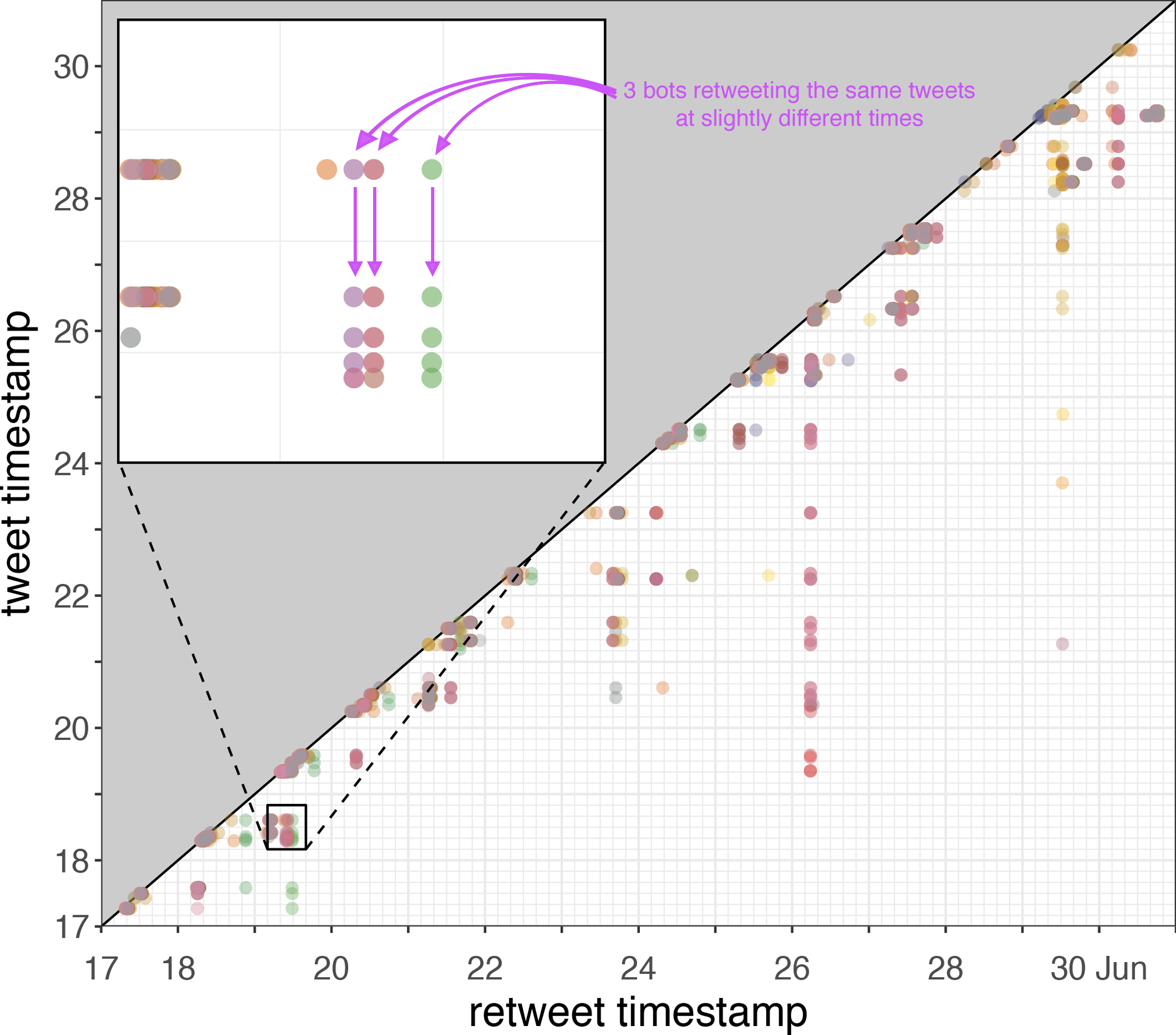}\caption{\textit{Cars} botnet.\label{fig:RTT-botnet-cars}}\end{subfigure}\begin{subfigure}[b]{0.24\textwidth}\centering\includegraphics[width=\textwidth]{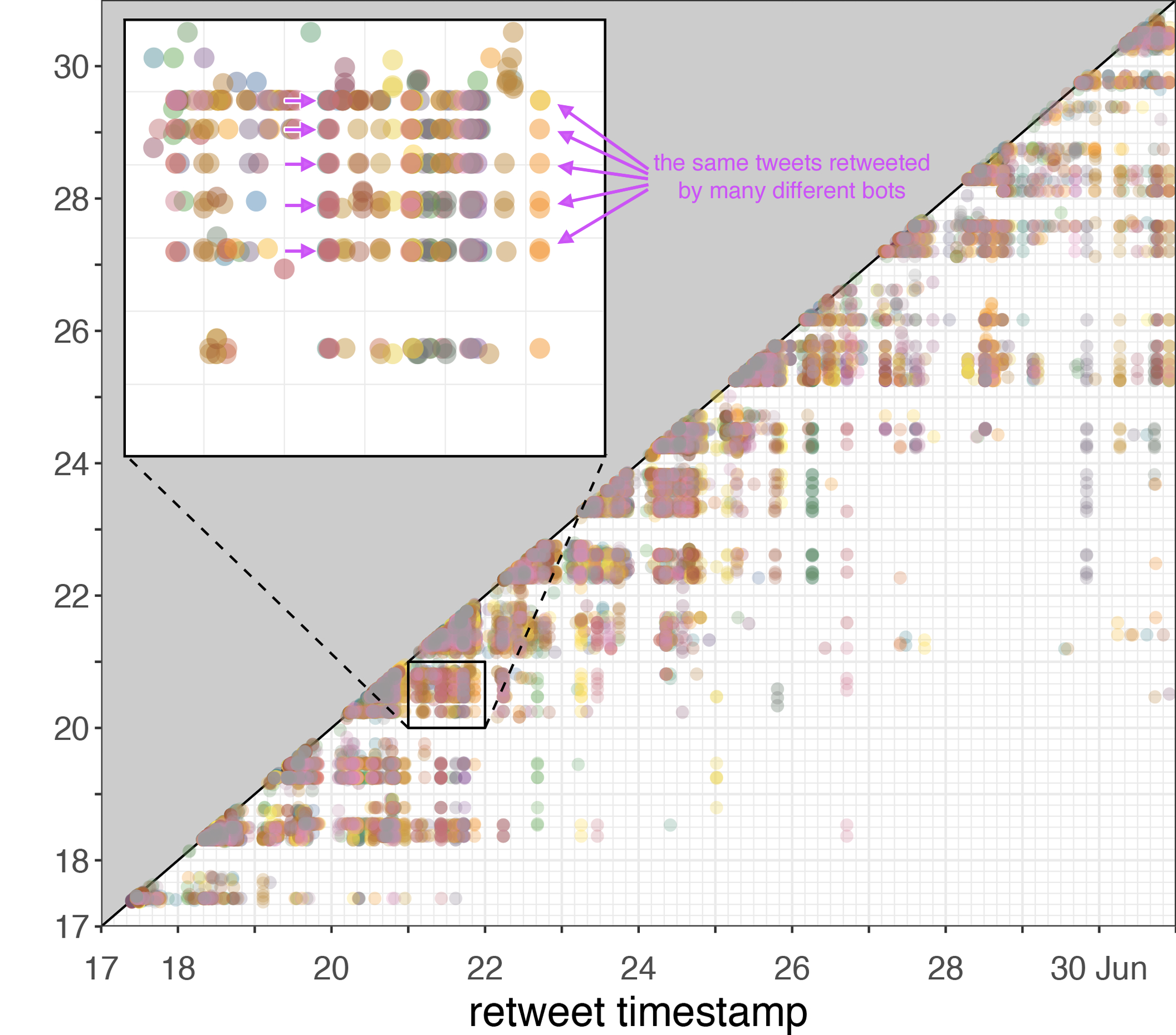}\caption{\textit{Singer} botnet.\label{fig:RTT-botnet-singer}}\end{subfigure}\caption{\textsc{\textmd{RTT}} plots of the 2 botnets discovered with \textsc{\textmd{RTbust}}.\label{fig:RTT-botnets}}
\end{figure}

\begin{figure}[t]
    \centering    \includegraphics[width=0.9\columnwidth]{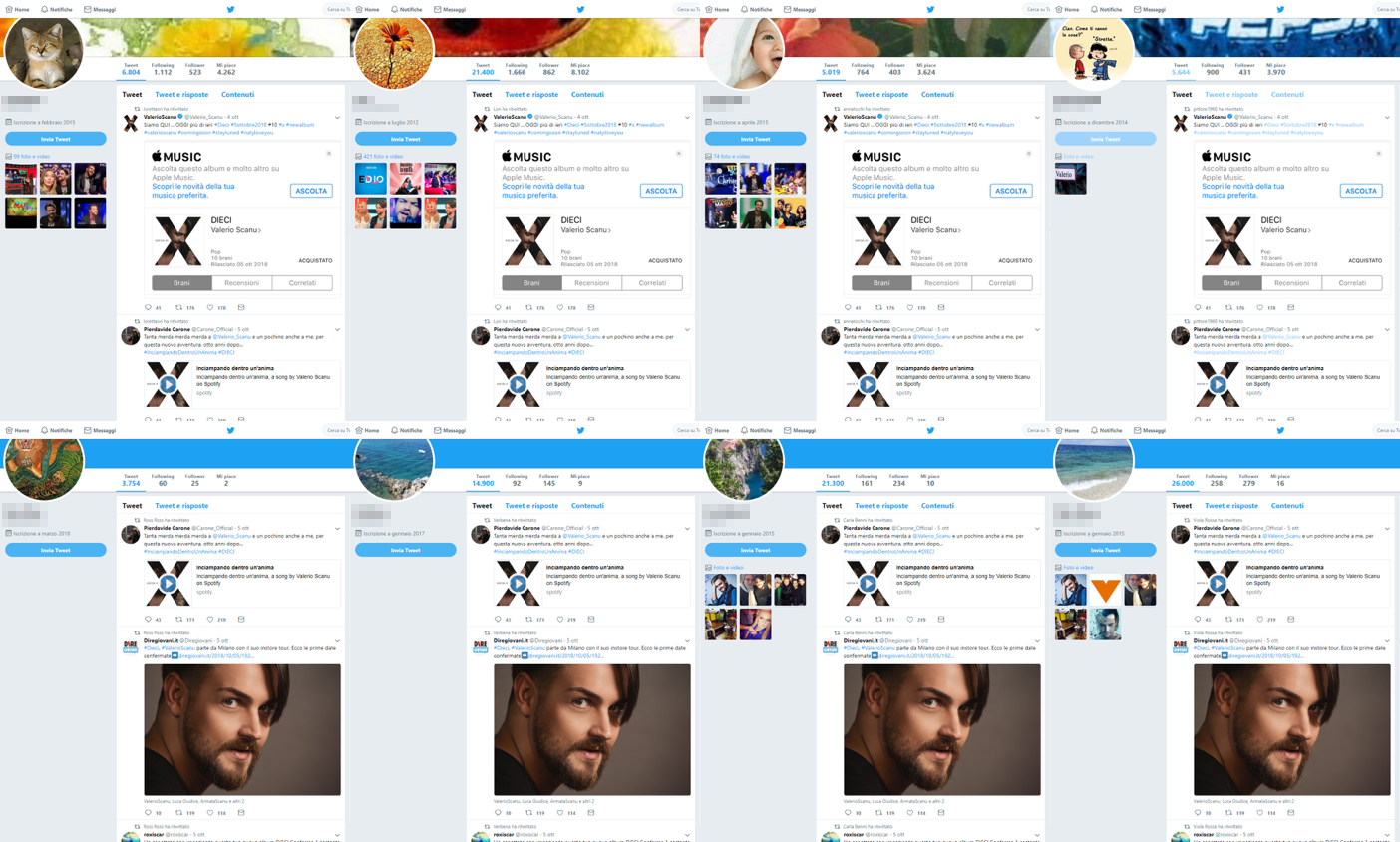}	\caption{A subset of accounts from the \textit{singer} botnet.\label{fig:singer-bots}}
\end{figure}

\textbf{Qualitative evaluation of bot detection.} Given the good results achieved by \textsc{RTbust} (VAE) at spotting the annotated bots in our dataset, we now analyze also those non-annotated accounts that have been labeled as bots by the same technique. This manual and qualitative validation might serve as an additional evaluation of the outcomes of \textsc{RTbust}.

In particular, \textsc{RTbust} found 2 notable clusters of non-annotated accounts. The smallest of such clusters counts 44 accounts. A manual inspection of these accounts revealed that they actually belong to a small botnet with some peculiar characteristics. First and foremost, they almost only retweet 3 accounts (\textit{@peugeotitalia}, \textit{@citroenitalia} and \textit{@motorionline}). Thus, this botnet has a specific focus on \textit{cars}. Moreover, they only post retweets or images, no account of the botnet has a real profile picture, and they are loosely synchronized, meaning that they tend to retweet the same tweets but at different moments in time. In order to get further insights into their behavior, we produced a combined \textsc{RTT} visualization of all the accounts of the botnet, as shown in Figure~\ref{fig:RTT-botnet-cars}. Each point in figure is a retweet made by one of the botnet's account and colors represent different bots. Interestingly, clear patterns emerge in some portions of the \textsc{RTT} plot, as seen in the inset of Figure~\ref{fig:RTT-botnet-cars}. Such patterns testify the synchronized retweeting activity of the botnet.

The other notable cluster of bot accounts found by \textsc{RTbust} is composed of almost 300 accounts that almost exclusively retweet 2 accounts (\textit{@Valerio\_Scanu} and \textit{@ArmataScanu}). Similarly to the previous smaller botnet, also this larger one seems to retweet with a specific focus -- that is, publicizing tweets related to the Italian pop singer \textit{Valerio Scanu}\footnote{\scriptsize\url{https://en.wikipedia.org/wiki/Valerio\_Scanu}}. Figure~\ref{fig:RTT-botnet-singer} shows the \textsc{RTT} plot of this botnet, which exhibits a mixture of the triangular and waterfall patterns that we described in Section~\ref{sec:exploration}. Figure~\ref{fig:singer-bots} also displays the appearance of the profiles of some of these accounts. As seen, some accounts of the botnet are tightly synchronized, while the botnet as a whole is loosely synchronized.

This qualitative evaluation revealed that, not only did \textsc{RTbust} achieve good results on the annotated accounts, but it also allowed to discover 2 previously unknown active retweeting botnets. 
\makeatletter{}\section{Discussion}
\label{sec:discussion}
\textbf{Visualizing suspicious behaviors.} In this work we provided several contributions. The first of such contributions is the development of the \textsc{RTT} visualization. In figures~\ref{fig:RTT-plots} and~\ref{fig:RTT-botnets} we showed that \textsc{RTT} plots represent a useful tool for analyzing the retweeting behavior of an account, or of a group of accounts. Human analysts can gain valuable insights into the behaviors of suspicious accounts by leveraging \textsc{RTT} plots. The possibility to leverage \textsc{RTT} plots might be particularly valuable in future data annotation tasks, considering the current difficulties faced by human annotators in correctly labeling social bots~\cite{Cresci2017}. Another favorable application scenario is related to the banning of bots from OSNs. Indeed, despite the recent advances in machine learning-based detectors, manual verification of accounts to assess their degree of automation is still carried out by OSN administrators~\cite{ferrara2016}. To this end, \textsc{RTT} plots might contribute to speed-up the process and to reduce possible human mistakes.

Notably, a few previous works also proposed some simple visualizations with the goal of highlighting suspicious behaviors in OSNs~\cite{giatsoglou2015retweeting,jiang2016inferring,jiang2016}. However, previous visualizations for spotting suspicious behaviors require much more data than that needed by \textsc{RTT} plots. For example, the visualizations used in~\cite{jiang2016} are based on the full social graph of considered accounts. Similarly, the visualizations proposed in~\cite{giatsoglou2015retweeting} require building retweet threads/cascades, which in turn require friend/followers information of each retweeter. Conversely, with \textsc{RTT} plots we can spot suspicious retweeting behaviors with as little information as retweet-tweet timestamps.

\textbf{Generalizability and robustness.} Our second, and largest, contribution is the development of the \textsc{Retweet-Buster} (\textsc{RTbust}) bot detection technique. \textsc{RTbust} is an unsupervised technique that is capable of automatically spotting meaningful patterns in retweet data. Because of this feature, \textsc{RTbust} is potentially capable of detecting retweeting bots exhibiting a behavior that has not been witnessed before. We believe this to be an important feature, considering that we still lack a ``standard'' and ``well-agreed'' definition of what a social bot is~\cite{maus2017typology,yang2019arming} -- and consequently, of its expected behaviors~\cite{almaatouq2014twitter}. Hence, the capability to spot a broad set of different behaviors, surely comes in handy. Moreover, it has been now largely demonstrated that social bots do \textit{evolve} to escape detection techniques~\cite{Cresci2017,cresci2018proaction}. Thus, the generalizability of \textsc{RTbust} and its robustness to evasion is a much desirable feature because it allows us to better withstand the next evolution of social bots.

\textbf{Explainability.} Model and decision explainability has now become one of the major practical and ethical concerns around AI~\cite{guidotti2018survey}. To this regard, one of the possible drawbacks of \textsc{RTbust} lies in the difficulty to ``explain'' the reasons for labeling an account as bot or legitimate. This is largely due to the difficulty in interpreting the latent features used for clustering, which in turn depend on the adoption of the ``black-box'' variational autoencoder for unsupervised feature extraction. This issue is not peculiar of our technique, but it is rather a well-known limitation of all deep learning techniques~\cite{guidotti2018survey}. Here, to mitigate this issue we again propose to resort to \textsc{RTT} plots. As demonstrated in Figure~\ref{fig:RTT-botnets}, our visualization can be profitably used also \textit{after} the application of \textsc{RTbust}, with the goal of understanding the characteristics of those accounts that have been labeled as bots. 
\makeatletter{}\section{Conclusions}
\label{sec:conc}
In this work, we investigated patterns of retweeting activity on Twitter, with the specific goal of detecting malicious retweeting bots. To this end, our work provides several contributions.

Firstly, we proposed a novel visualization technique called \textsc{ReTweet-Tweet} (\textsc{RTT}). As thoroughly shown, \textsc{RTT} plots are an effective and efficient mean to gain valuable insights into the retweeting behaviors of Twitter accounts. By leveraging \textsc{RTT} plots we analyzed the ``normal'' retweeting behavior of legitimate users, and we uncovered 3 suspicious behaviors that are caused by automated retweeting -- and thus, that are representative of bot activities. Furthermore, we discussed how \textsc{RTT} plots can empower human analysts when manually annotating a dataset comprising social bots, as well as OSN administrators looking for evidence of automation when deciding about banning accounts from social platforms. Finally, we highlighted that \textsc{RTT} plots can also be used to explain decisions of black-box bot detectors, thus contributing towards explainable and interpretable AI.

Next, we designed an unsupervised group-analysis technique, called \textsc{Retweet-Buster} (\textsc{RTbust}), for detecting retweeting social bots. The core of \textsc{RTbust} is an LSTM variational autoencoder that we trained to extract a minimum number of highly informative latent features from the retweet time series of each account. The decision about an account (whether it is bot or legitimate) is based on the outcome of a hierarchical density-based clustering algorithm. In detail, accounts belonging to large clusters are labeled as bots, while unclustered accounts are labeled as legitimate. We compared different implementations of \textsc{RTbust} with baselines and state-of-the-art social bot detection techniques. \textsc{RTbust} outperformed all competitors achieving $F1 = 0.87$, in contrast with $F1 \leq 0.76$ of other techniques. By applying \textsc{RTbust} to a large dataset of retweets, we also discovered 2 previously unknown active botnets comprising hundreds of accounts.

For future work we plan to improve the decisions taken by \textsc{RTbust}. In fact, here we adopted a rather naive solution revolving around accounts being clustered or not. However, we can augment the final decision step of \textsc{RTbust} by considering additional information, such as the hierarchy of clusters produced by the clustering algorithm and other internal and external clustering validation measures. In this way, we could prune some clusters or some accounts belonging to a cluster, thus improving \textit{precision}. Similarly, we could expand some clusters by adding borderline accounts, thus possibly also improving \textit{recall}, which currently is the bottleneck of \textsc{RTbust}. Other than improving \textsc{RTbust}, we also plan to apply it \textit{in the wild} for discovering and analyzing active botnets.

\bibliographystyle{ACM-Reference-Format}
\bibliography{references}

\end{document}